\documentclass[twocolumn,epjc3]{svjour3}         

\RequirePackage[T1]{fontenc}
\smartqed  
\usepackage[utf8]{inputenc}
\RequirePackage{graphicx}
\RequirePackage{mathptmx}      
\RequirePackage{flushend}
\RequirePackage[numbers,sort&compress]{natbib}
\usepackage{graphics}
\usepackage{xurl}
\usepackage{float}
\usepackage{subcaption}
\usepackage[switch,columnwise]{lineno}
\usepackage{testhyphens}
\usepackage{hyphenat}
\usepackage{mathptmx}
\usepackage{color}
\usepackage{graphicx} 
\usepackage{microtype}

\input{my_ds_macro.def}

\newcommand{\APC}{APC, Université de Paris Cit\'e, CNRS, Astroparticule et Cosmologie, Paris F-75013, France}
\newcommand{\AQLNGS}{INFN Laboratori Nazionali del Gran Sasso, Assergi (AQ) 67100, Italy}
\newcommand{\AQGSSI}{Gran Sasso Science Institute, L'Aquila 67100, Italy}

\newcommand{\CAUniELE}{Department of Electrical and Electronic Engineering, Universit\`a degli Studi di Cagliari, Cagliari 09042, Italy}
\newcommand{\CAUniPHY}{Physics Department, Universit\`a degli Studi di Cagliari, Cagliari 09042, Italy}
\newcommand{\CAINFN}{INFN Cagliari, Cagliari 09042, Italy}
\newcommand{\Carleton}{Department of Physics, Carleton University, Ottawa, ON K1S 5B6, Canada}

\newcommand{\CIEMAT}{CIEMAT, Centro de Investigaciones Energ\'eticas, Medioambientales y Tecnol\'ogicas, Madrid 28040, Spain}

\newcommand{\Hawaii}{Department of Physics and Astronomy, University of Hawai'i, Honolulu, HI 96822, USA}

\newcommand{\NAINFN}{INFN Napoli, Napoli 80126, Italy}

\newcommand{\Oxford}{University of Oxford, Oxford OX1 2JD, United Kingdom}

\newcommand{\Princeton}{Physics Department, Princeton University, Princeton, NJ 08544, USA}

\newcommand{\SRNL}{Savannah River National Laboratory, Jackson, SC 29831, USA}

\newcommand{\UCDavis}{Department of Physics, University of California, Davis, CA 95616, USA}

\newcommand{\Zaragoza}{Centro de Astropartículas y Física de Altas Energías (CAPA), Universidad de Zaragoza, Zaragoza, 50009, Spain}
\newcommand{\AstroCeNT}{AstroCeNT, Nicolaus Copernicus Astronomical Center of the Polish Academy of Sciences, Warsaw, 00-614, Poland}
\newcommand{\MPK}{Max-Planck-Institut f\"ur Kernphysik, Saupfercheckweg 1, 69117 Heidelberg, Germany}
\newcommand{\PadovaUni}{Department of Physics, Universit\`a degli Studi di Padova, Padova, Italy}

\newcounter{autoinst}
\newcommand{\definst}[1]{%
  \stepcounter{autoinst}%
  \expandafter\xdef\csname num#1\endcsname{\theautoinst}%
}

\definst{CAINFN}     
\definst{CAUniPHY}   
\definst{CIEMAT}     
\definst{SRNL} 
\definst{Zaragoza}   
\definst{AQLNGS}     
\definst{UCDavis}    
\definst{Oxford}    
\definst{Carleton}   
\definst{AstroCeNT}  
\definst{APC}        
\definst{AQGSSI}     
\definst{NAINFN}     
\definst{CAUniELE}   
\definst{MPK}        
\definst{Hawaii}     
\definst{Princeton}  
\definst{PadovaUni}  

\graphicspath{{figs/}}

\journalname{Eur. Phys. J. C}

\usepackage{url}
\usepackage[
  breaklinks,
 colorlinks=true,    
 linkcolor=blue,     
 citecolor=blue,     
 filecolor=blue,  
 urlcolor=blue,      
 final=true
]{hyperref}
\begin{document}
\runningpagewiselinenumbers
\title{Measurement of the $^{39}$Ar specific activity in atmospheric argon with DArT at the Canfranc Underground Laboratory
}






\author{
P.~Agnes\textsuperscript{\numAQLNGS,\numAQGSSI},
H.~O.~Back\textsuperscript{\numSRNL},
S.~Bharat\textsuperscript{\numZaragoza},
W.~M.~Bonivento\textsuperscript{\numCAINFN}%
\thanks{Corresponding author1:
\href{mailto:walter.bonivento@ca.infn.it}
{walter.bonivento@ca.infn.it}}
,
M.~Boulay\textsuperscript{\numCarleton},
M.~Caboni\textsuperscript{\numCAINFN},
N.~Canci\textsuperscript{\numNAINFN},
N.~Cargioli\textsuperscript{\numCAINFN},
P.~Castello\textsuperscript{\numCAINFN,\numCAUniELE},
S.~Cebrián\textsuperscript{\numZaragoza},
L.~Consiglio\textsuperscript{\numAQLNGS},
R.~Crampton\textsuperscript{\numCarleton},
D.~Diaz~Mairena\textsuperscript{\numCIEMAT},
D.~Franco\textsuperscript{\numAPC},
I.~Fuente-Ortega\textsuperscript{\numCIEMAT},
F.~Gabriele\textsuperscript{\numCAINFN},
D.~Gahan\textsuperscript{\numAQLNGS},
C.~Galbiati\textsuperscript{\numPrinceton},
P.~Garcia~Abia\textsuperscript{\numCIEMAT},
T.~Hessel\textsuperscript{\numAPC},
S.~Horikawa\textsuperscript{\numAQLNGS,\numHawaii},
G.~Korga\textsuperscript{\numOxford},
A.~Ianni\textsuperscript{\numAQLNGS},
M.~Ku\'{z}niak\textsuperscript{\numAstroCeNT},
L.~Luzzi\textsuperscript{\numUCDavis},
A.~Marasciulli\textsuperscript{\numAQLNGS},
M.~Martínez\textsuperscript{\numZaragoza},
P.~Organtini\textsuperscript{\numAQLNGS,\numPrinceton},
A.~Ortiz de Solórzano\textsuperscript{\numZaragoza},
R.~Pavarani\textsuperscript{\numCAINFN,\numPadovaUni},
P.~Pegoraro\textsuperscript{\numCAINFN,\numCAUniELE},
V.~Pesudo~Fortes\textsuperscript{\numCIEMAT}
\thanks{Corresponding author2:
\href{mailto:vicente.pesudo@ciemat.es}
{vicente.pesudo@ciemat.es}},
M.~Razeti\textsuperscript{\numCAINFN},
O.~U.~Rehman\textsuperscript{\numCAINFN},
L.~Romero\textsuperscript{\numCIEMAT},
D.~Sablone\textsuperscript{\numAQLNGS},
E.~S\'{a}nchez~Garc\'{i}a\textsuperscript{\numMPK},
R.~Santorelli\textsuperscript{\numCIEMAT},
C.~Seoane\textsuperscript{\numZaragoza},
J.~Sosiak\textsuperscript{\numCarleton},
S.~Sulis\textsuperscript{\numCAINFN,\numCAUniELE},
C.~Tierney\textsuperscript{\numCarleton},
S.~Tullio\textsuperscript{\numCAINFN,\numCAUniPHY},
G.~Vera~D\'{\i}az\textsuperscript{\numCIEMAT}
}

\institute{
$^{\numCAINFN}$ \CAINFN. \\
$^{\numCAUniPHY}$ \CAUniPHY. \\
$^{\numCIEMAT}$ \CIEMAT. \\
$^{\numSRNL}$ \SRNL. \\
$^{\numZaragoza}$ \Zaragoza. \\
$^{\numAQLNGS}$ \AQLNGS. \\
$^{\numUCDavis}$ \UCDavis. \\
$^{\numOxford}$ \Oxford. \\
$^{\numCarleton}$ \Carleton. \\
$^{\numAstroCeNT}$ \AstroCeNT. \\
$^{\numAPC}$ \APC. \\
$^{\numAQGSSI}$ \AQGSSI. \\
$^{\numNAINFN}$ \NAINFN. \\
$^{\numCAUniELE}$ \CAUniELE. \\
$^{\numMPK}$ \MPK \\
$^{\numHawaii}$ \Hawaii. \\
$^{\numPrinceton}$ \Princeton. \\
$^{\numPadovaUni}$ \PadovaUni.
}

\date{\today}

\maketitle

\abstract{We report the measurement of the $^{39}$Ar specific activity of atmospheric argon using DArT, a low-back\-ground single-phase liquid-argon detector read out by cryogenic silicon photomultipliers, at the Canfranc Underground Laboratory (LSC) in Spain. 
 DArT was first filled with atmospheric argon and then with a sample of underground argon of known radioactivity from the DarkSide-50 experiment.
  The underground argon serves as a background reference in this measurement, making the result robust because we do not rely on background simulations. 
We measure the $^{39}$Ar specific activity with both a cut-and-count method and a binned maximum-likelihood fit, yielding consistent results. We obtain 
$a_{\text{AAr}}= 0.955 \pm 0.008~\text{Bq}/\text{kg}$. 
 This result agrees with previous measurements from other experiments and provides the most precise determination of the $^{39}$Ar specific activity in atmospheric argon to date. It also  validates DArT as a key component of the  DArTInArDM experiment at LSC, which aims to measure the $^{39}$Ar activity of the argon 
extracted from deep underground wells in Colorado (USA) for the \DSk\ and LEGEND-1000 experiments at Laboratori Nazionali del Gran Sasso in Italy. 
}

\section{Introduction}
\label{sec:intro}

Liquid argon is a widely used target for astroparticle physics and, in particular, for direct WIMP dark matter searches because of its scalability, high scintillation yield, and powerful pulse-shape discrimination between electron and nuclear recoils. However, atmospheric argon (AAr) contains the cosmogenic isotope \ce{^39Ar}, a $\beta$-emitter with an activity of about 1~Bq/kg, which constitutes an intrinsic source of electron-recoil background.

A precise determination of the \ce{^39Ar} activity in atmospheric argon is important for various reasons. 

Indeed,  atmospheric argon is the reference against which all other argon sources are compared, especially argon extracted from underground reservoirs. 

Moreover, for future experiments involving hundreds~\cite{Aalseth:2018gq} or even thousands of tonnes~\cite{DUNE:2020txw, DUNE:2024wvj} of liquid argon, a small uncertainty in the specific activity translates into a very large uncertainty in the total number of radioactive decays. It can influence the design of the trigger and data-acquisition systems, the background model, the optimization of pulse-shape discrimination, and, ultimately, the experiment's projected sensitivity.

Precise measurements of the \ce{^39Ar} specific activity and  half-life~\cite{DEAP:2025shk} allow the determination of the isotopic abundance and the comparison with theoretical models of \ce{^39Ar} production by cosmic rays~\cite{PhysRevC.100.024608, Bhattacharya:2025emx}.

The relevance of \ce{^39Ar} is not limited to particle or nuclear physics. It is also widely used to date groundwater, ice, and ocean water over timescales ranging from several decades to roughly 1,000 y~\cite{Loosli1983Ar39Dating, Ritterbusch2014Ar39ATTA}. The uncertainty in atmospheric activity directly propagates into the calculated age, and the reference uncertainty is especially important for relatively young samples.

Liquid argon detectors offer a particularly suitable environment for this measurement, i.e., direct radioactivity counting with a very low-background device.  Indeed, the radioactive isotope is uniformly distributed throughout the active volume, and the detector itself acts as both the source and the detection medium. This makes it possible to study many decays with high efficiency, avoiding uncertainties associated with external sources, sample preparation, and geometrical acceptance.



Previous measurements of the specific \ce{^39Ar} activity in atmospheric liquid argon, ${a}_{\text{AAr}}$,  are summarized in Tab.~\ref{tab:spec_meas} and are consistent with each other.

\begin{table}[ht!]
	\centering
    	\caption{Measurements of specific activity of atmospheric argon, ${a}_{\text{AAr}}$, from other experiments. }
        \footnotesize
	\begin{tabular}{ll}
		\hline
		experiment       & ${a}_{\text{AAr}}$ [Bq/kg]  \\
        \hline 
        DEAP-3600~\cite{DEAP:2023wri} &  $0.964 \pm 0.001_{\text{stat.}}\pm 0.024_{\text{syst}}$  \\
        WARP~\cite{WARP:2006nsa} &  1.01$\pm$ 0.02 $_{\text{stat.}}\pm 0.08 _{\text{syst}}$   \\
        ArDM~\cite{ArDM:2017ndf} &  $0.95\pm 0.05$ \\
        \hline 
	\end{tabular}
	\label{tab:spec_meas}
\end{table}

In this paper, we present a new measurement using previously characterized low-radioactivity underground argon to constrain the background and reduce dependence on background modeling.
We used the 
\DArT\ detector, a low-background device with a two-SiPM readout, immersed in a cryostat with a liquid-nitrogen bath at the Canfranc Underground Laboratory (LSC) in Spain.

DArT's design requirements come from the DArTInArDM  project~\cite{DarkSide-20k:2020qfz}, which is being commissioned at the same underground laboratory. Its main goal is the \ArThirtyNine\ radioactivity assaying of the argon extracted from underground wells with the Urania plant 
in Colorado~\cite{Aalseth:2018gq, DarkSide-20k:2024inx, GlobalArgonDarkMatter:2024wtv}, and from the purification plant Aria~\cite{DarkSide-20k:2021nia, DarkSide-20k:2023grj} in Italy, for use in the \DSk\ and LEGEND-1000 experiments at Laboratori Nazionali del Gran Sasso in Italy (LNGS).

A limited amount of underground argon (UAr), about 150~kg, was already extracted from the same well in Colorado, processed at Fermilab (IL), USA, and used for the \DSf\ experiment at LNGS~\cite{DarkSide:2018kuk, DarkSide:2018bpj}. We measured its \ce{^39Ar} radioactivity with the \DSf\ detector to be ($0.73\pm 0.11$)~mBq/kg, i.e., a factor $1400\pm 200$ smaller than that of atmospheric argon (AAr). We use UAr as the reference for our measurements, as described below.

In Sect.~\ref{sec:detector} we describe the \DArT\ detector, including the mechanical structure, light detection, readout electronics, the cryogenic test setup and its operation.  In Sect.~\ref{sec:sim} we present the detector response model. In Sect.~\ref{sec:Uar} we introduce the UAr used in this measurement. In Sect.~\ref{sec:response}, we describe the detector response for AAr and UAr data, and  a measurement with an external  
$^{137}$Cs source. 
In Sect.~\ref{sec:purity} we present the measurement of liquid argon purity over time. In Sect.~\ref{sec:activity} we present the \ArThirtyNine\ specific activity measurement. 

\section{The DArT detector}
\label{sec:detector}
DArT is a compact single-phase liquid-argon detector instrumented with two cryogenic silicon photomultipliers. It was developed as a prototype and test-bench detector for the DArTInArDM program to validate the mechanical design, gas-system operation, cryogenic SiPM readout, data-acquisition system, and analysis chain in liquid argon. The detector was assembled and first tested on the surface at CIEMAT in Madrid, Spain, before being transported to the LSC. The data presented in this paper were acquired underground at LSC, with DArT operated in the dedicated cryogenic test setup. The following subsections describe DArT's main hardware components: the mechanical structure of the copper vessel and internal acrylic assembly, the light-collection system, the cryogenic SiPM readout, and the dedicated cryogenic setup used for the AAr and UAr measurements.

\subsection{Mechanical structure}

\begin{figure}[ht!]
\centering
\includegraphics[width=0.5\columnwidth]{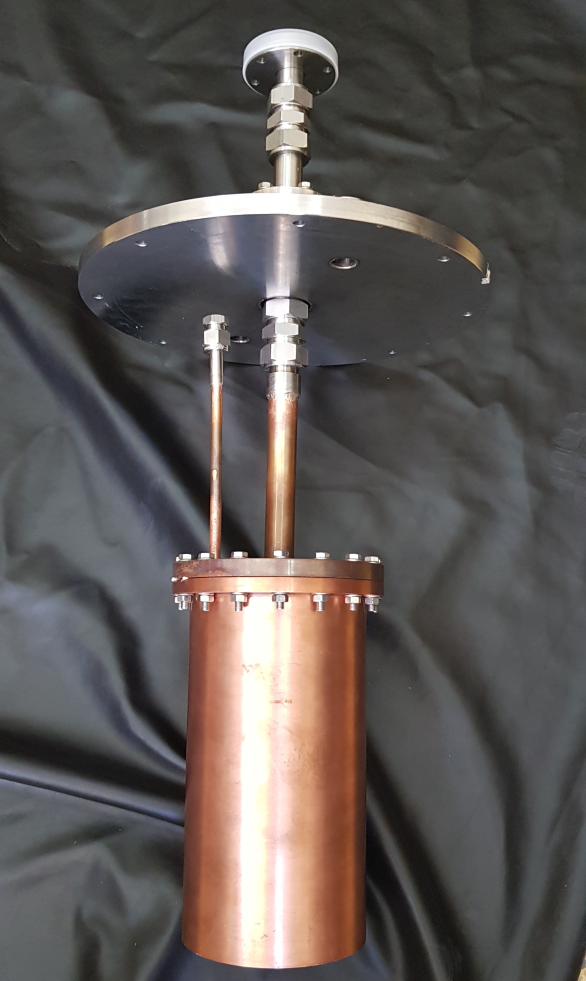}
\caption[]{The DArT copper vessel, connection pipes, and the cryostat top flange.}
\label{fig:vessel}
\end{figure}

The DArT detector is housed inside a cylindrical vessel made of ultra-pure Oxygen-Free High-Conductivity (OFHC) copper, as shown in Fig.~\ref{fig:vessel}. The cylinder has an open top, which is closed by a flange. The wider central pipe serves as both the gas connection and cable access, while the thinner pipe is blind.
 The flange is closed using 16 stainless steel (SS) bolts and four  Belleville washers per bolt. We achieve tightness by placing indium wire in grooves carved into the copper. To improve gas tightness at low temperatures over repeated cooling cycles, we reinforce the copper lid and flange with two SS rings.

\begin{figure}[ht!]
\centering
\includegraphics[angle=-90,width=0.5\columnwidth]{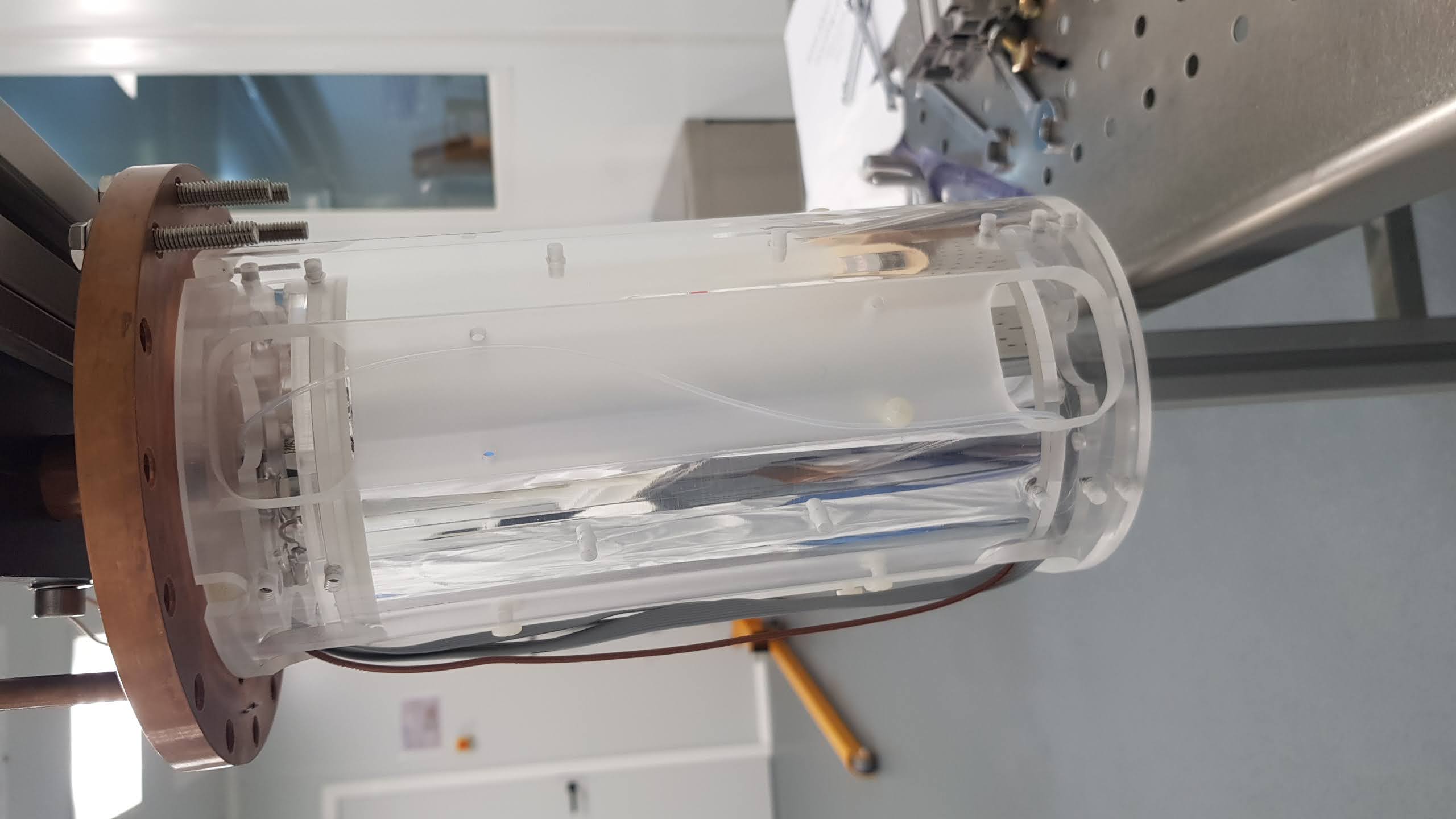}
\caption{The inside of the DArT detector after removing the cylindrical copper vessel.}
\label{fig:acrylic}
\end{figure}

Inside the copper vessel, there are two nested acrylic cylinders. The external acrylic cylinder holds the DArT detector and is bolted to the top flange of the copper vessel. This outer cylinder, shown in Fig.~\ref{fig:acrylic}, provides mechanical support, sensor mounting points, and cable routing. 
The top and bottom caps of the outer cylinder host platinum resistance PT100 sensors that serve as temperature sensors, level meters, and heat sources to facilitate liquid argon filling and boil-off.

The inner cylinder, $\sim$169~mm high and with an inner diameter of $\sim$85.5~mm, together with its top and bottom caps, defines DArT's active volume. 
The inner surface of the cylinder and the caps are sandblasted and coated with 200 $\mathrm{\mu}$g/cm$\mathrm{^2}$ of wavelength shifter 1,1,4,4-tetraphenyl-1,3-butadiene (TPB). An aluminum-based Mylar reflector is sandwiched between the inner and outer cylinders and completely covers the inner cylinder's external surface. Two disc-shaped ESR films are placed on the external sides of the caps. Each reflector has a centered rectangular opening, $1.2 \times 0.8~\mathrm{cm}^2$, through which photons reach the SiPM. The SiPMs are housed in two acrylic structures attached to the external cylinder.

\subsection{Light detection and readout  electronics}
\label{sec:electronics}



\begin{figure}[ht!]
\centering
\includegraphics[width=0.7\columnwidth]{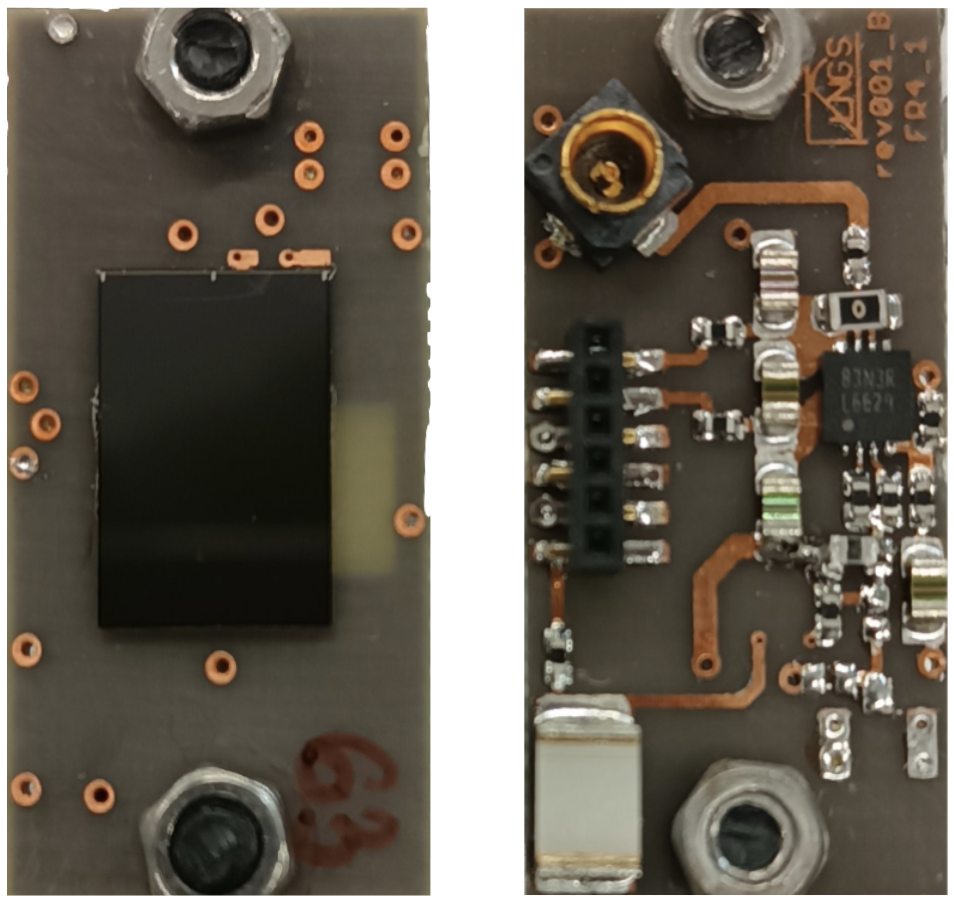}
\caption{The DArTEye readout board housing: on the front side, the SiPM; on the back side, the front-end electronics.}
\label{fig:board}
\end{figure}

A custom readout board based on discrete components, DArTEye,  housing one SiPM, was designed and is shown in Fig.~\ref{fig:board}. The board substrate is made of Arlon 55NT, a radiopure material which matches the coefficient of thermal expansion of silicon.  The light sensors are $11.7\times 7.9$~mm$^2$ NUV-HD-Cryo SiPMs with 30~$\mu$m micro-cell pitch, of the same type used by DarkSide-20k, designed by the Fondazione Bruno Kessler (FBK), with production wafers manufactured by LFoundry (AV), Italy~\cite{Gola2019, DarkSide-20k:2025avf, Organtini:2020bga}. 

The front-end processing is based on the same transim\-pedance amplifier (TIA) design used in \DSk~\cite{DIncecco2:2018hy,   Kugathasan:2020xry}, where it is used to read out a tile, with minimal modifications to
improve the amplifier stability with the readout of a single SiPM.

A Keithley 2450 Graphical SourceMeter supplied the SiPM bias voltage of 31~V, and the  Keysight E3643A power supply provided the low voltage (+1.6~V, -3.4~V) to the front-end electronics.
 SiPM signals are transmitted through micro-miniature MMCX cables to a 16-channel, 14-bit CAEN V1730SB digitizer, operating at a
sampling rate of 500~MS/s, corresponding to a sampling
interval of 2~ns.
 The acquisition length was  10~\(\mu\)s, enough to detect most of the scintillation light of the LAr and perform pulse shape discrimination.
We set the trigger time to sample 1500, and used the time before the trigger to measure the signal baseline and noise level.
We set the trigger threshold to 3.66~mV above the baseline, and during data acquisition, we configured the two channels in OR.
A MIDAS-based software platform collects digitized data, stores it locally, and then transfers the files to grid storage for subsequent reconstruction and analysis.

We characterized a batch of 20~DArTEye boards by measuring I-V curves in liquid nitrogen. 
All boards exhibit similar behavior, with a breakdown voltage of approximately 27~V.  We selected two of them for the measurements described in this paper. 
For all the measurements reported here, we always operated the SiPMs at an overvoltage of 4~V.

\subsection{The cryogenic test setup}
\label{sec:cryosetup}

\begin{figure}[ht!]
\centering
\includegraphics[width=0.7\columnwidth]{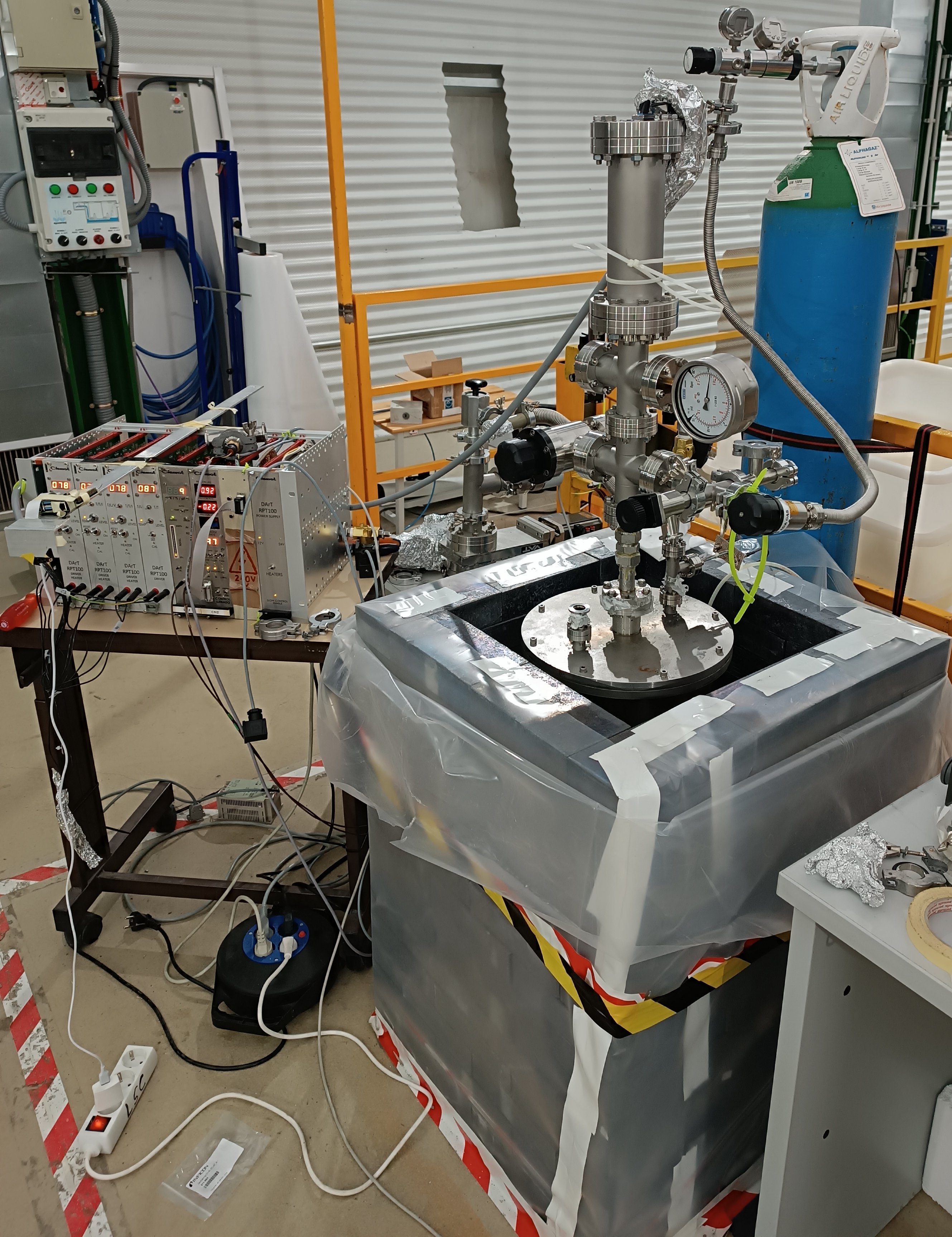}
\caption[]{DArT in the cryogenic test setup. The top cover with lead bricks was not mounted when this picture was taken.}
\label{fig:test_setup}
\end{figure}

We developed the cryogenic test setup to test DArT independently of ArDM. In the final setup, DArT will be immersed in the ArDM liquid-argon bath, which will provide both stable cryogenic conditions and an active veto. For the measurements reported here, we reproduced this environment with a dedicated liquid-nitrogen cryostat. 
To suppress background from external radiation, we surrounded the cryostat with a 10~cm-thick lead brick shield, as shown in Fig.~\ref{fig:test_setup}.
Since the temperature of liquid nitrogen at atmospheric pressure is below the freezing point of argon, the cryostat was operated at $\sim$2.1 bar to raise the nitrogen temperature and keep the argon in DArT in the liquid phase. This pressure-controlled operation was therefore a key requirement of the test setup.
DArT was filled by introducing argon gas into the detector volume, which condensed on the cold copper walls of the chamber and pipes. This approach exploits copper's high thermal conductivity, which efficiently transfers heat released during argon cooling and condensation to the surrounding cryogenic bath. The argon gas was passed through a hot getter before entering the detector, reducing impurities that could quench the scintillation light. During filling from 50~L bottles pressurized to 200~bar for the AAr and 30~bar for the UAr, the DArT vessel was maintained above atmospheric pressure to promote condensation and prevent back-diffusion of contaminants. The average filling rate achieved in the test setup was about 12.8~mL/min of liquid argon, corresponding to an extracted heat power of approximately 80~W.
The cryogenic system included an automated liquid-nitrogen filling and pressure-control system. The setup included a PT100 temperature sensor, a pressure transducer, a capacitive level meter, and two electromagnetic valves controlling the liquid-nitrogen inlet and the cryostat vent. An Arduino Mega-based electronic module read out and controlled these elements, allowing the cryostat level and pressure to be regulated during operation. This control system was necessary to maintain stable thermal conditions over long data-taking periods and to safely compensate for nitrogen boil-off.
In addition to providing the required thermal environment, the test setup enabled evaluation of DArT's gas tightness under realistic cryogenic conditions. Nitrogen intake from the surrounding bath would quench the slow component of liquid-argon scintillation, thereby providing a sensitive diagnostic of possible leaks or contamination. The long-term stability of the scintillation time profile over several-day runs confirms that the cryogenic and gas-handling system provided suitable conditions for measuring the \ce{^39Ar} activity.

\section{Detector response model}
\label{sec:sim}



We developed the DArT detector response model to account for geometrical effects, TPB wavelength shifting, reflector properties, and the light-collection efficiency of the  SiPMs. This is particularly important for DArT, where the compact geometry and the presence of acrylic between the TPB-coated surfaces and the reflectors can significantly affect photon transport, the amount of light reaching the sensors, and the reconstructed energy scale. 
The simulation chain models DArT's response to both \ce{^39Ar} $\beta$-decays and calibration data. It generates energy deposits in the active LAr volume, propagates VUV scintillation photons, applies TPB wavelength shifting, transports visible photons to the SiPMs, generates signal shapes, and digitizes waveforms to reproduce the measured electronic response.

The $^{39}$Ar $\beta$-decay spectrum used in the simulation comes from Ref.~\cite{MOUGEOT2023111018} and includes atomic screening and exchange corrections.

\subsection{Optical response model}


	
	

VUV photons reaching the TPB are absorbed and re-emitted isotropically into the visible band. We then track the visible photons until they are absorbed by the SiPMs or other materials.
The optical properties of material interfaces are those of the default version used by the \DSf\  experiment.

We treated the SiPM quantum efficiency and ESR reflectivity as effective optical parameters, constrained by the comparison between data and simulation in the \ce{^137Cs} calibration run, described in Sect.~\ref{sec:calibration}. The values used in the nominal simulation are consistent with independent measurements~\cite{WinNT} and are subsequently varied within their uncertainties to assess the corresponding systematic contribution.

Only about 5\%  of the visible photons impinge on the SiPMs. The rest do not contribute to the signal because they are channeled in the acrylic via total internal reflection at the acrylic/TPB and TPB/LAr interfaces.
Unfortunately, the acrylic structure between the TPB and the reflector acted as an unwanted light guide, trapping visible photons.

\subsection{Electronic response}

We used a Python-based toolkit to produce a digitized waveform similar to the raw data, using the time-profile input from the optical simulation of visible photons impinging on the SiPMs. To simulate the effect of quantum efficiency (QE), we adopt a simplified approach that relies on a uniform random number generator and neglects its wavelength dependence. We then convolve the photoelectron time profile with the Single Photoelectron (SPE) response function and the noise spectrum, both determined from data, as described in Sect.~\ref{sec:response}.
Tab.~\ref{tab:electsim} shows the parameters used to simulate the electronic response, derived from test-bench measurements or other test setups (see, e.g., Ref.~\cite{Boulay:2021njr}).

\begin{table}[ht!]
	\centering
    	\caption{Parameters of the electronics simulation. }
	\begin{tabular}{ll}
		\hline
		parameter       &  value  \\
        \hline
        noise RMS    & 1.2 mV  \\
        dark count rate & 10$^{-3}$ Hz/mm$^2$ \\ 
        after-pulse probability &    10\%  \\
        direct cross-talk  & 10\%  \\
        \hline
	
	\end{tabular}

	\label{tab:electsim}
\end{table}

At cryogenic temperatures, the dark-count contribution is negligible for the event selection, whereas after-pulsing and crosstalk can modify the reconstructed charge and are therefore included in the nominal waveform simulation and varied in the systematic studies.


\section{Underground argon}
\label{sec:Uar}

The underground argon sample used in this work consists of argon from the DarkSide-50 experiment. Because DarkSide-50 previously measured its residual \ce{^39Ar} activity, this sample provides a suitable low-activity reference for comparison with atmospheric argon.

About 100~kg of UAr from the DarkSide-50 decommissioning were stored underground at LNGS. Three 50~L bottles, each corresponding to 2.5~kg of argon, were filled and shipped to LSC. 

According to Ref.~\cite{DarkSide:2015cqb}, the UAr at the time of the measurement contained ($0.73\pm 0.11$)~mBq/kg of $\rm ^{39}$Ar and  ($2.05\pm 0.13$)~mBq/kg  of  $\rm ^{85}$Kr.

The UAr measurements in \DSf\ were taken from April to July 2015, and those described in this paper were taken during May 2024, i.e., 9~y later.
Given the \ce{^85Kr} half-life of ($10.739\pm 0.014$)~y and the \ce{^39Ar} half-life of (302$\pm$ 10)~y~\cite{DEAP:2025shk}, the  \ce{^85Kr} and \ce{^39Ar} activities had decreased by 2024  to:
\begin{eqnarray}
    {a}^{^{85}\text{Kr}}_{\text{UAr}} & = & (1.15\pm  0.07)~\text{mBq/kg} \nonumber \\ 
    {a}^{^{39}\text{Ar}}_{\text{UAr}} & = &  (0.71\pm 0.11)~\text{mBq/kg} \label{eq:uar}
\end{eqnarray}


Before operating DArT with this UAr sample, a gas analysis using a  Thermo Scientific inductively coupled plasma mass spectrometer (ICP-MS) was performed to assess whether it had undergone significant atmospheric contamination during storage, transport, or handling at LNGS after the \DSf\ run, and to investigate the presumed excess of \ce{^85Kr}. This is relevant because an air intake after the \DSf\ run would have introduced atmospheric argon, which could increase the \ce{^39Ar} content of the sample and bias the AAr/UAr comparison. 
A \(^{39}\)Ar activity of \(0.1\) mBq/kg in the UAr would produce a relative bias of approximately \(0.01\%\) on the measured AAr activity. Such a contamination would correspond to approximately 1\% air infiltration by volume,
i.e., 0.8\% nitrogen and 0.2\% oxygen, which were excluded by the ICP-MS analysis, which measured at least an order of magnitude smaller contamination for both species. Our conclusion is therefore that no air infiltration at a level that could affect the present AAr/UAr comparison occurred in the UAr sample after the \DSf\ run. 


The same ICP-MS analysis also allowed us to assess the possible $^{85}$Kr contribution in the AAr reference sample used for the present $^{39}$Ar measurement, comparing it with its concentration in the UAr sample.

The presence of krypton in the UAr \DSf\ sample is believed to result from air infiltration during extraction in 2014 (i.e., before the \DSf\ run) and from the limited effectiveness of chemical impurity removal in the distillation system. ICP-MS analysis detected all stable krypton isotopes in the UAr sample, with relative abundances consistent with those of atmospheric krypton.
%

The isotope $^{85}$Kr is man-made, produced in nuclear fission and released by nuclear fuel reprocessing plants and nuclear weapon tests. The relative isotopic abundance of $^{85}\text{Kr}/^\text{nat}\text{Kr}$ was estimated  to  be $2 \times 10^{-11}~\text{mol/mol}$ in 2014~\cite{Lindemann:2013kna}.

Assuming that krypton entered the UAr from the atmosphere in 2014 with the isotopic ratio quoted above, we use the absolute krypton concentration inferred from the ICP-MS calibration with strontium isotopes to estimate the \(^{85}\mathrm{Kr}\) specific activity. After accounting for radioactive decay between the relevant dates, the estimate is consistent with the DarkSide-50 measurement.
Strontium isotopes were used as calibration references because their masses are close to those of the krypton isotopes.


The gas used for the \ce{^39Ar} specific activity measurement is argon Alphagaz\texttrademark~2 (1~ppm impurity level, as declared by the manufacturer). With this gas, ICP-MS found the stable krypton concentration to be approximately 10$^{-4}$ times that in the UAr sample. 
The ratio of stable krypton concentrations can therefore be used to estimate the relative \(^{85}\mathrm{Kr}\) contributions, provided that differences in the isotopic ratios due to the origin and storage history of the two samples are accounted for. 
Given the four-order-of-magnitude lower krypton concentration in AAr, plausible variations in these ratios do not affect the conclusion that its \(^{85}\mathrm{Kr}\) contribution is negligible.

\section{Detector response with AAr and UAr}
\label{sec:response}

\begin{figure}[ht!]
	\centering
\includegraphics[width=\columnwidth]{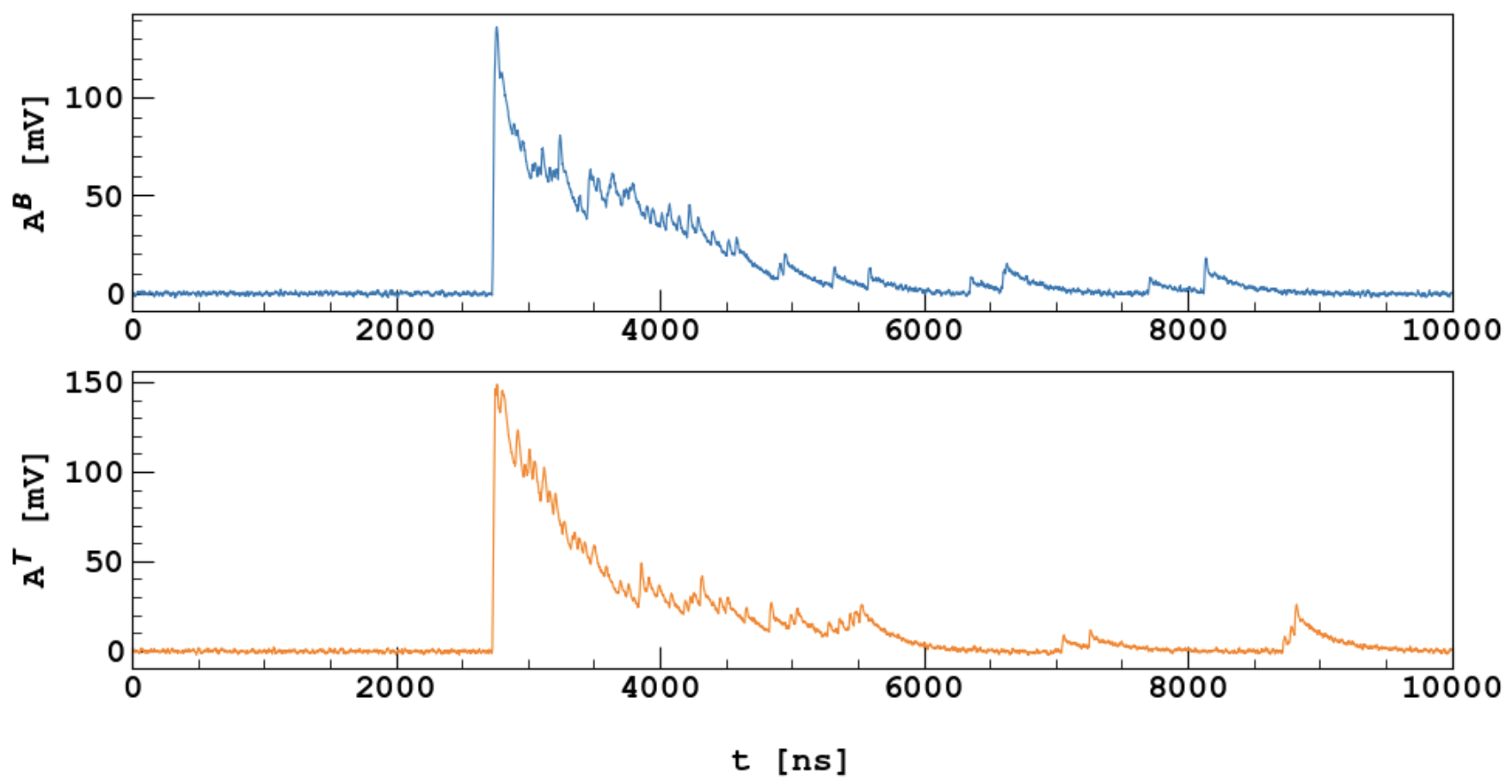}
	\caption{Signal amplitude ($A$) vs. time ($t$) for the top and bottom channels in a single event.}
	\label{fig:wfsamp}
\end{figure}

Fig.~\ref{fig:wfsamp} shows a typical waveform from the run data. 
The signal shape can be understood as a superposition of many photon signals, with time constants for singlet ($\tau_{s}=2-6$~ns) and triplet ($\tau_{t}=1.1-1.6~\mu$s) light emission from liquid argon scintillation, convolved with the SiPM response.
The different regions of interest for reconstruction are the pre-trigger up to 2.25~$\mu$s, the prompt region, starting from the trigger time and lasting 400~ns, whose time integral defines the variable $Q_\text{prompt}$, and the full region, lasting 5~$\mu$s, whose time integral defines the variable $Q_\text{tot}$ or total charge. 

\begin{figure}[ht!]
\centering
\includegraphics[width=\columnwidth]{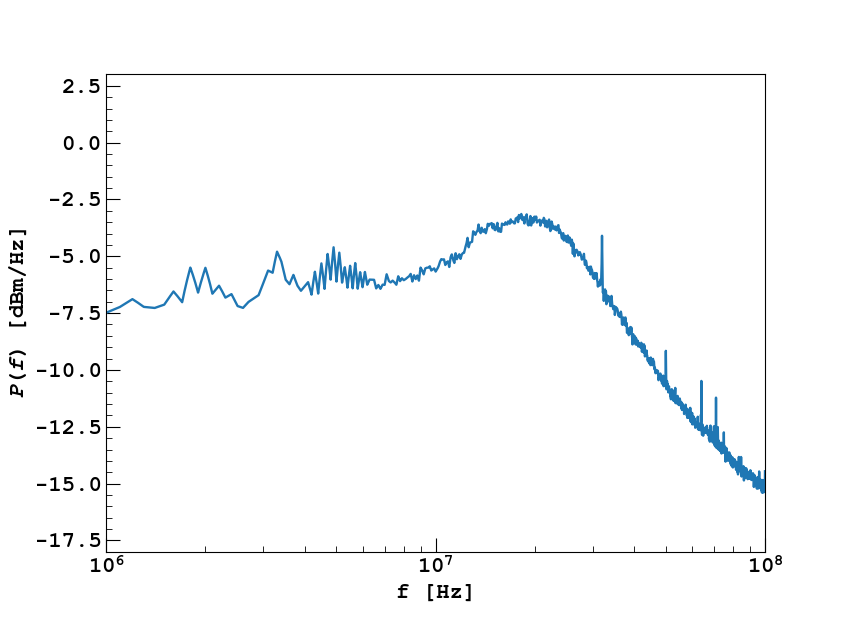}
\caption{Output noise power spectrum ($P(f)$) versus frequency ($f$).}
\label{fig:noise_spectrum}
\centering
\end{figure}

The output noise power spectrum, \(P(f)\), with $f$ the frequency, was determined from events in which one channel failed to pass the trigger threshold, and averaged over 200 events, as:
\begin{equation}
P(f) = \log_{10} |\text{FFT}_{\text{norm}}({A(t)})|^{2},
\end{equation}
where  \text{FFT}$_{\text{norm}}$  is the Fast Fourier Transform of the time-dependent waveform $A(t)$,
and is shown in Fig.~\ref{fig:noise_spectrum}.
The output noise power spectrum peaks at $\sim$20~MHz.

\begin{figure}[ht!]
\centering
\includegraphics[width=\columnwidth]{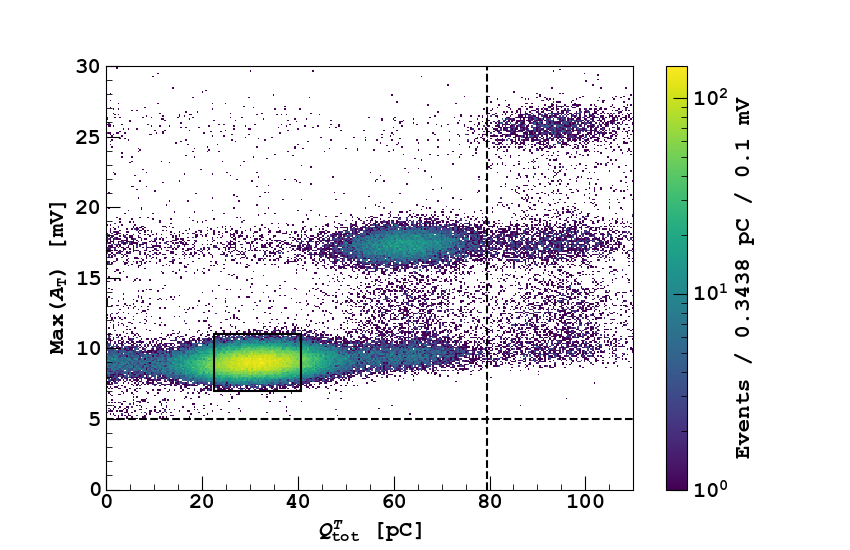}
\caption{Maximum signal amplitude (Max($A_\text{T}$))   vs  charge $Q^\text{T}_\text{tot}$~(pC)  in the top channel. The black box corresponds to the selected events for single-photoelectron charge.  }
\label{fig:bidi_PH}
\centering
\end{figure}

 Fig.~\ref{fig:bidi_PH}  shows, for the top channel, the maximum signal amplitude in the full integration window vs. total charge  $Q^\text{T}_\text{tot}$  distribution, by selecting events with a threshold requirement of 5~mV on both channels.
A clear pattern emerges with peaks corresponding to one, two, and more photoelectrons. 

Saturation in the electronic chain occurs at a signal amplitude of 1.2~V, so we implemented it in the simulation. The number of events below 200~PE with at least one saturated channel is negligible in both the data and the simulation. In the range of the $\alpha$-decays, instead, all events have saturated signals.

\begin{figure}[ht!]
	\centering
	\includegraphics[width=\columnwidth]{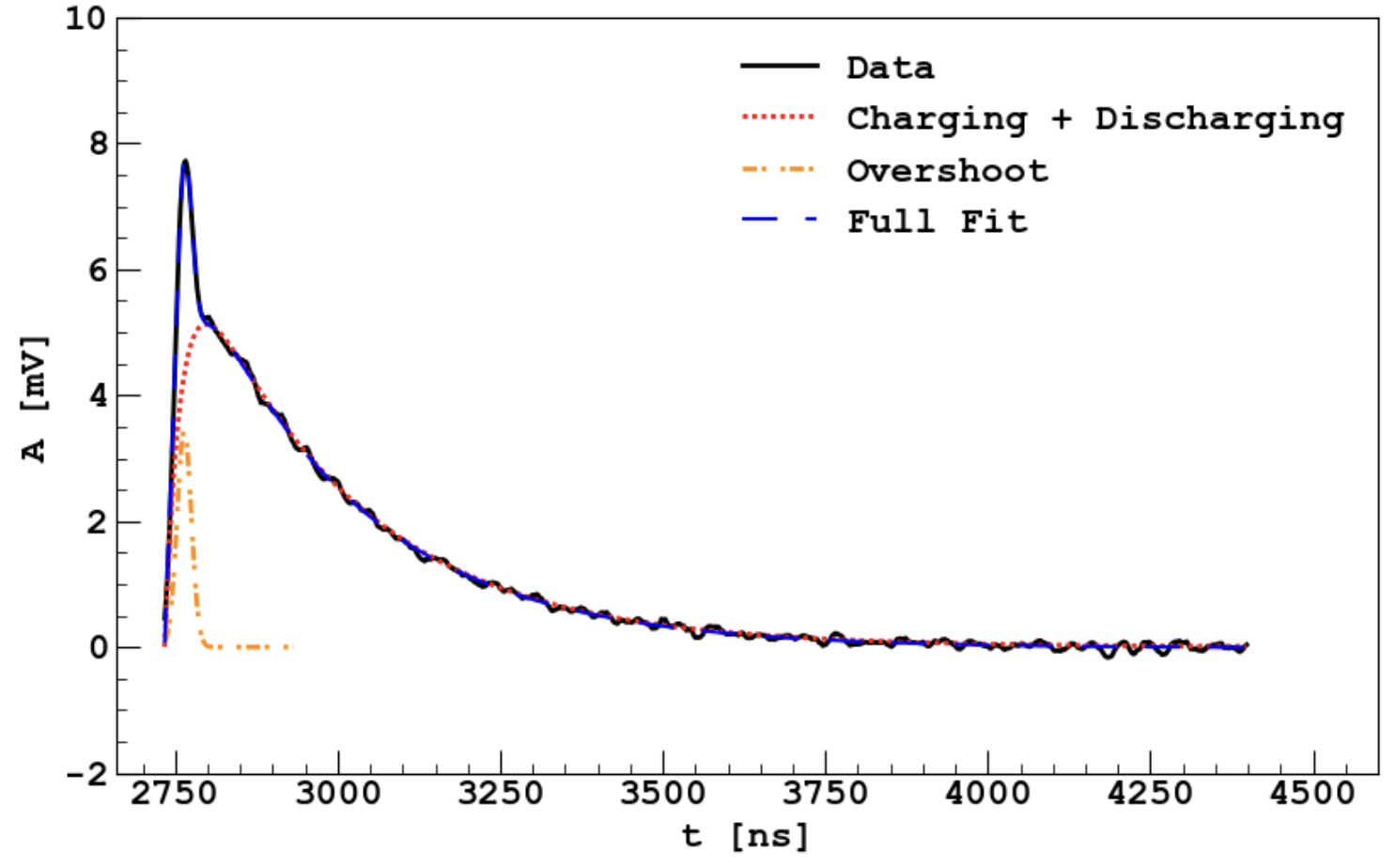}
	\caption{Signal amplitude ($A$) vs. time ($t$) for events corresponding to a single-photoelectron charge, as selected from Fig.~\ref{fig:bidi_PH}. Average of 300 events. Overlaid is the fitted response function, defined in Eq.~\ref{eq:spe}. }
	\label{fig:spefit}
\end{figure}

To extract the pulse shape corresponding to a single photoelectron, we used the following cuts:  $Q^\text{T}_\text{tot}$  between 22.5~pC and 40.5~pC and maximum signal amplitude between 7 and 10.5~mV.
The waveform shown in Fig.~\ref{fig:spefit} is the average of 300 selected events.
The  waveform is fitted with the following  function:
\begin{equation}
\text{SPE}(t) = A_{\text{SPE}}\left(e^{\frac{(t_0 - t)}{t_2}} - e^{\frac{(t_0 - t)}{t_1}}\right) + A_1 e^{-\frac{(t - m)^2}{2\sigma^2}}
\label{eq:spe}
\end{equation}
where \(t_0\) is a constant time offset.  
The third term describes the overshoot at the beginning of the pulse and is modeled with a Gaussian distribution with mean m and standard deviation $\sigma$. This overshoot is a feature already observed in other SiPMs at liquid-nitrogen temperatures; see, e.g., Ref.~\cite{Otono:2006zz}.
The fit results are  \(t_{2} = 242 \pm 10\)~ns, \(t_{1}  = 21 \pm 4\)~ns, \(\sigma  = 4.780 \pm 0.008\)~ns, and \(m  = 18 \pm 4\)~ns 
with a  \(\chi^2/n_\text{dof}\)  of 1.1.
We used the mean fit values as simulation input parameters.

\begin{figure}[ht!]
\centering
\includegraphics[width=\columnwidth]{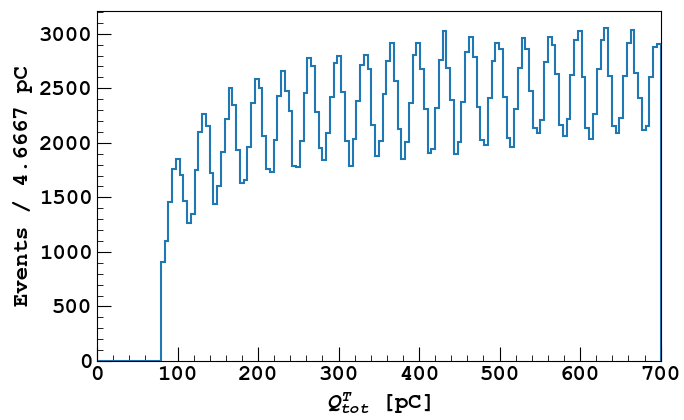}
\caption{
$Q^\text{T}_\text{tot}$~(pC)   distribution of the top channel in AAr data at low charge. }
\label{fig:PE_data_top}
\end{figure}
\begin{figure}[ht!]
	\centering	\includegraphics[width=1.1\columnwidth]{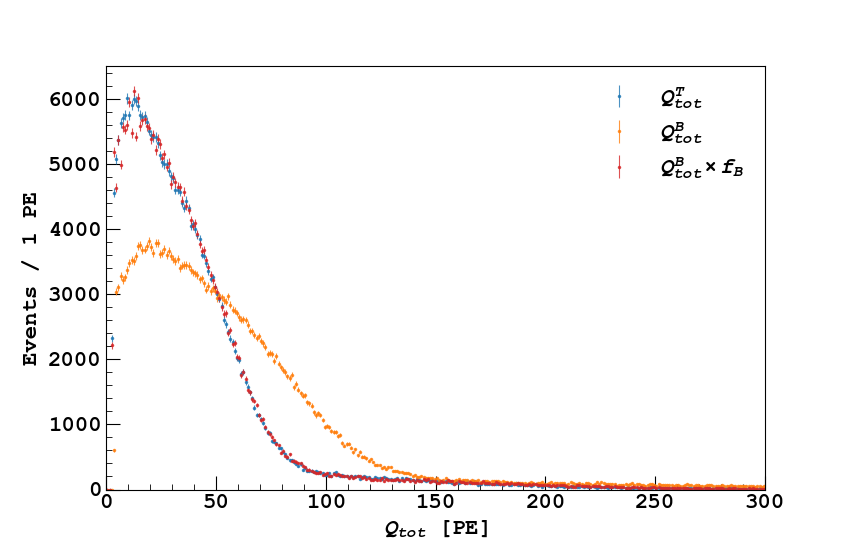}
	\caption{$Q^\text{T,B}_\text{tot}$(PE)  distribution for top (blue dots) and bottom channel, before (orange dots) and after (red dots) f$_{B}$ correction.}
	\label{fig:corrf}
\end{figure}

Fig.~\ref{fig:PE_data_top} 
shows the distribution of the charge response $Q^\text{T}_\text{tot}$~(pC)  for the top SiPM in AAr data for $Q^\text{T}_\text{tot} <$~700~pC, requiring that both channels measure at least 80~pC.
Gain calibration in PE for the $Q^\text{T,B}_\text{tot}$  distribution for the two channels was performed by fitting the first 20 peaks
with Gaussian functions. 
The measured peak separations were fitted with 
a second-order polynomial as:
\begin{equation}
    \Delta Q^\text{T,B}_\text{tot}=\mu_N - \mu_1 = a+b (N-1) +c (N-1)^2,
\label{eq:fitgain}
\end{equation} 
with $N$ the number of photoelectrons,  allowing a free intercept to accommodate residual deviations from the calibration model.
Tab.~\ref{tab:gaincalib} shows the fit results. Linearity is observed up to 20~PE (the non-linear term $c$ is very small). The fitted gain slopes in data and simulation agree to approximately 5\%. We also verified that gain calibration in the first and second halves of the run agrees within 0.4\% for both UAr and AAr. The agreement between UAr and AAr gain calibration is also within 0.5\%.
Using the fitted function of Eq.~\ref{eq:fitgain} we obtain 
\begin{table*}[ht!]
\footnotesize
	\centering
    	\caption{SiPM gain calibration based on the fitting procedure described in the text. The fit parameters correspond to the function of Eq.~\ref{eq:fitgain} and $\sigma_2$ is the width of the peak corresponding to two photoelectrons.}
	\begin{tabular}{lccccc}
    \hline
    channel & $\mu_1$ (pC) & $a$ (pC) & $b$ (pC/PE) & $c$ (pC/PE$^2$) & $\sigma_2$ (pC) \\
    \hline
    \multicolumn{6}{l}{data AAr} \\
    \hline
    top    & $30.71 \pm 0.03$ & $-0.60 \pm 0.07$ & $31.58 \pm 0.02$ & $-0.004 \pm 0.001$ & $7.68 \pm 0.01$ \\
    bottom & $30.01 \pm 0.18$ & $ 0.33 \pm 0.23$ & $31.38 \pm 0.06$ & $0.000 \pm 0.003$ & $7.70 \pm 0.01$ \\
    \hline
    \multicolumn{6}{l}{simulation} \\
    \hline
    top    & $32.35 \pm 0.26$ & $1.18 \pm 0.28$ & $33.11 \pm 0.07$ & $0.002 \pm 0.003$ & $7.86 \pm 0.11$ \\
    bottom & $31.60 \pm 0.32$ & $2.34 \pm 0.35$ & $33.01 \pm 0.08$ & $0.006 \pm 0.004$ & $8.37 \pm 0.24$ \\
    \hline
\end{tabular}

	\label{tab:gaincalib}
\end{table*}
the $Q^\text{T}_\text{tot}$(PE)  and $Q^\text{B}_\text{tot}$(PE)  distributions,  shown in Fig.~\ref{fig:corrf}.
The top channel has an average charge about 30\% lower than the bottom channel. 
Given that the simulation reproduces the SiPM calibration well, we attribute this to a light-collection effect. We confirmed this at the end of data collection, when we opened the detector. A visual inspection revealed a vertical misalignment between the top and bottom SiPMs inside the opening of the end face of the external acrylic cylinder, with a displacement of about  \DArTtopsipmdisplacement.
To account for this effect, we implemented an equivalent displacement in the simulation. 
From a maximum-likelihood fit, we determine the multiplicative correction factors to apply to the bottom SiPM to equalize its charge response to that of the top one: $f_{B}^D=0.648\pm 0.002$ in data and $f_{B}^S=0.646\pm 0.002$ in simulation.  Fig.~\ref{fig:corrf} also shows the scaled $Q^\text{B}_\text{tot}$(PE)  distribution in data. 

\begin{figure}[ht!]
\centering
\includegraphics[width=\columnwidth]{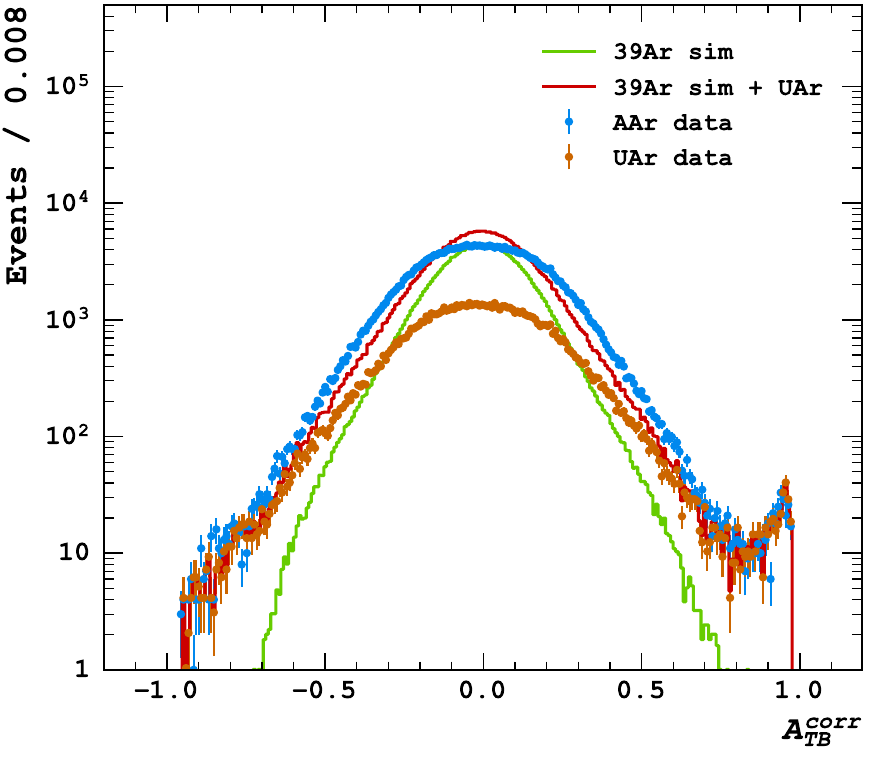}
\caption{Distribution of the corrected top--bottom asymmetry,
$A^{\mathrm{corr}}_{\mathrm{TB}}$, for events with
$Q_{\mathrm{tot}}<200\,\mathrm{PE}$.
Shown are AAr data (blue points), UAr data scaled to the
AAr live time (orange points), simulated ${}^{39}\mathrm{Ar}$
events (green solid line), and the sum of the simulation
and the scaled UAr spectrum (red solid line).
The simulated signal is normalized to the specific activity
reported in Eq.~\ref{eq:act_fin}.}
\label{fig:TBA_comp}
\end{figure}

Fig.~\ref{fig:TBA_comp} shows the corrected top-bottom asymmetry, defined as:
\begin{equation}
    A^{\mathrm{corr}}_{\mathrm{TB}} = \frac{Q_{\mathrm{tot}}^{T} - f^{D,S}_{B}\,Q_{\mathrm{tot}}^{B}}{Q_{\mathrm{tot}}^{T} + f^{D,S}_{B}\,Q_{\mathrm{tot}}^{B}},
    \label{eq:tba_corr}
\end{equation}
for AAr data, UAr data scaled to the
AAr live-time, simulated ${}^{39}\mathrm{Ar}$
events, and the sum of the simulation
and the scaled UAr spectrum.
The width of the latter distribution is smaller than that of the  AAr data distribution. 
Therefore, to estimate the impact of this residual
mismodelling on the signal efficiency, the simulated \ce{^39Ar} sample was
reweighted in bins of 
$A^{\mathrm{corr}}_{\mathrm{TB}}$.
We derived the weights from a control selection identical to the nominal one, except that we removed the per-channel $2.5~\mathrm{PE}$ threshold. We then used the reweighted simulation to recompute the nominal threshold selection efficiency.
The resulting change in efficiency was 0.14\%, which was assigned as the systematic uncertainty. 

 We also noticed a rise in the AAr and UAr distributions above 0.85 for $A^{\mathrm{corr}}_{\mathrm{TB}}$. The origin of this rise is not fully understood and is not observed in the simulation. We conservatively quote the time-normalized difference in the number of entries between AAr and UAr data in that region as a systematic uncertainty on the specific activity. 
\subsection{Energy scale calibration}
\label{sec:calibration}

To calibrate the absolute energy scale in the simulation, we compared data and simulation of a $^{137}$Cs source calibration run.

\begin{figure}[ht!]
\centering
\includegraphics[width=\columnwidth]{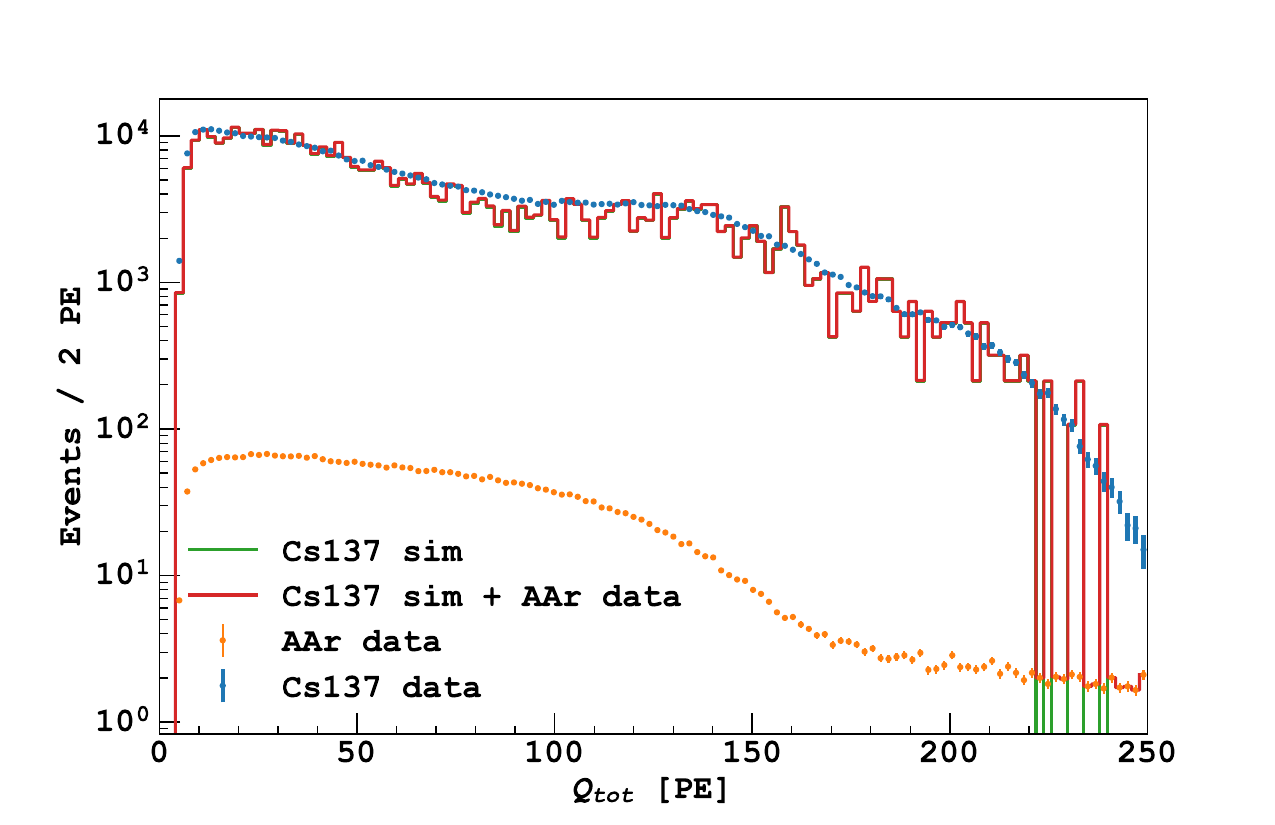}
\caption{$Q_{\text{tot}}$~(PE) distribution for a run acquired with a 27~kBq $^{137}$Cs source (black dots), AAr data (orange dots), and simulation of $^{137}$Cs added to it, normalized to the AAr run time. }
\label{fig:cesium}
\centering
\end{figure}

The total, $Q_{\text{tot}}$, and prompt, $Q_{\text{prompt}}$, charges are obtained by summing $Q^\text{T,B}_{\text{tot}}$ and $Q^\text{T,B}_{\text{prompt}}$, respectively, after applying the top-bottom correction factor $f_\text{B}$. A residual dimensionless scale factor, $S_\text{f}$, accounts for the remaining difference between the charge scale in data and simulation: 
\begin{eqnarray}
Q_{\text{tot}} & =& S_\text{f} (Q^\text{T}_{\text{tot}}+ f^\text{D,S}_{\text{B}} Q^\text{B}_{\text{tot}} ) = S_\text{f} Q^0_{\text{tot}} \\
Q_{\text{prompt}}&=&S_\text{f}(Q^\text{T}_{\text{prompt}}+f^\text{D,S} _{\text{B}} Q^\text{B}_{\text{prompt}}  ).
\end{eqnarray}
Fig.~\ref{fig:cesium} shows the $Q_\text{tot}$~(PE)   distribution for a run acquired with a 27~kBq $^{137}$Cs  source, AAr data, and simulation of $^{137}$Cs  added to it. We positioned the source at approximately mid-height in the chamber.

 The overall calibration factor $S_\text{f}$  was obtained by performing a binned maximum likelihood fit of the simulated $Q_{\text{tot}}$ distribution to that of the data, by constraining the total number of events to the source activity and run time, and we obtained $S_\text{f} = 1.034\pm 0.003$ (with $S_\text{f} = 1$ for the data).

We measure the unweighted
light yield (LY) as:
\begin{equation}
\mathrm{LY}  =  \frac{Q^\text{T}_{\text{tot}}+Q^\text{B}_{\text{tot}}}{E} =  (0.39\pm 0.03)~\text{PE/keV}, 
\end{equation}
where $E$ is the deposited energy expressed in keV.
From the \ArThirtyNine\ simulation, the LY turns out to be approximately independent of $z$ and $r$. 
This relatively low LY value is due to 
the photon-trapping between the acrylic structure between the TPB and the
reflector, as described in Sect.~\ref{sec:sim}.

\subsection{Pulse shape discrimination}
\label{sec:psd}

Scintillation in liquid argon consists of triplet and singlet emission with different time constants, as discussed in Sect.~\ref{sec:response}.  Electron recoil events and nuclear recoil events are associated with different combinations of these light emissions, enabling powerful particle identification via pulse-shape discrimination (PSD)~\cite{Amaudruz:2017ekt, DarkSide:2015cqb, DEAP:2021axq}. 
We define the PSD variable, $f_{\text{prompt}}$, as:
\begin{equation}
f_{\text{prompt}}=\frac{Q_{\text{prompt}}}{Q_{\text{tot}}}. 
\end{equation}

\begin{figure}[ht!]
\centering
\includegraphics[width=\columnwidth]{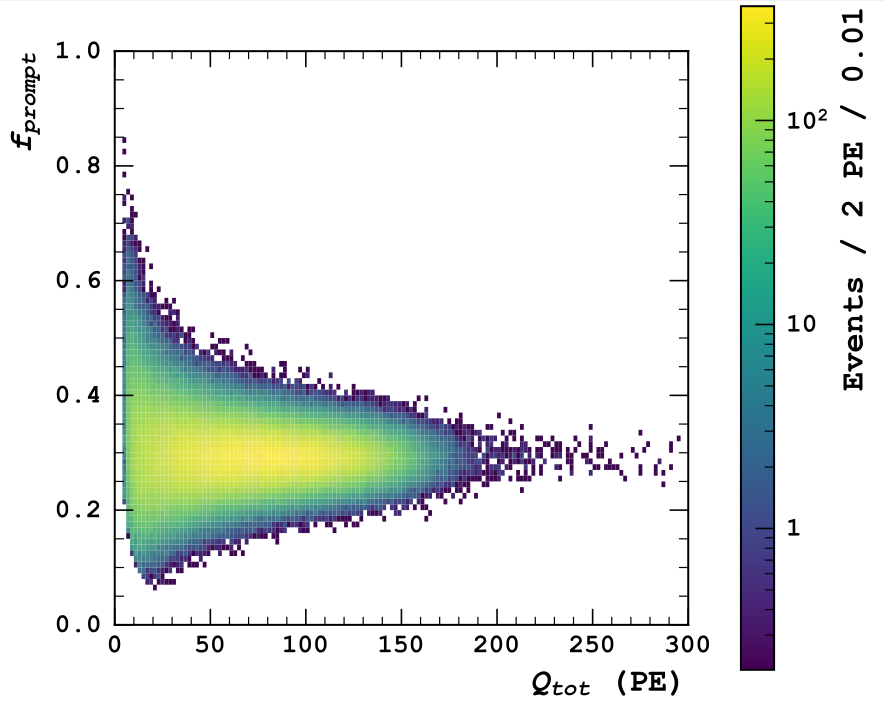}
\caption{$f_{\text{prompt}}$ vs $Q_{\text{tot}}(\text{PE)}$ for simulated \ArThirtyNine\ decays.}
\label{fig:PSD_sim}
\centering
\end{figure}

Fig.~\ref{fig:PSD_sim} shows  $f_{\text{prompt}}$ vs $Q_{\text{tot}}(\text{PE)}$ in simulated data for \ArThirtyNine\ decays.

\begin{figure}[ht!]
\centering
\includegraphics[width=\columnwidth]{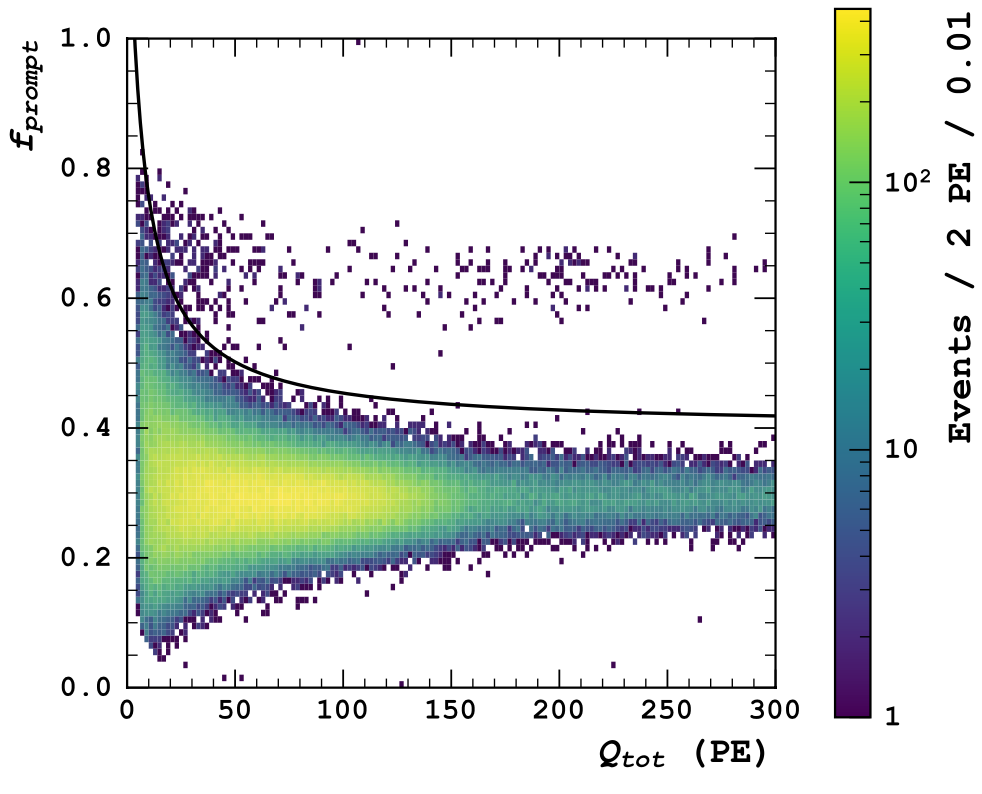}
\caption{$f_{\text{prompt}}$ vs $Q_{\text{tot}}(\text{PE)}$ in AAr data. The cut to remove nuclear-recoil events is shown. The selected events for the \ce{^39Ar} specific activity measurement are those below the black line.}
\label{fig:PSD_AAr}
\centering
\end{figure}

At large $Q_{\text{tot}}(\text{PE)}$, the mean value of $f_{\text{prompt}}$ is around 0.3, consistent with the expected prompt-charge fraction for electron-recoil scintillation events in liquid argon.
Fig.~\ref{fig:PSD_AAr} 
shows $f_{\text{prompt}}$  vs $Q_{\text{tot}}(\text{PE)}$ in AAr  data for $Q_{\text{tot}}(\text{PE)}<300$.
Another set of events clusters at $f_{\text{prompt}} \sim 0.65$, consistent with heavily ionizing events such as $\alpha$-interactions and neutron-induced nuclear recoils. 

To reject nuclear-recoil backgrounds, we apply the two-dimensional selection shown in Fig.~\ref{fig:PSD_AAr}, retaining high signal efficiency despite the poor \(f_{\mathrm{prompt}}\) resolution at low charge.

A few events, 4 in AAr and 2 in UAr, are observed in the data at small \(f_{\mathrm{prompt}}\) values, below 0.1, in a region where no events are present in the simulation, and are related to  $A^{\mathrm{corr}}_{\mathrm{TB}}$ values above 0.9. Their contribution to the systematic uncertainty on the \ce{^39Ar} specific activity is negligible.  


\section{Purity monitoring}
\label{sec:purity}

DArT's gas-tightness is essential for its final use in the ArDM setup. In this paper, we monitor potential nitrogen and oxygen infiltrations by measuring the slow component of the scintillation~\cite{Bonivento:2024qpn}. The quenching of the slow component also reduces $Q_{\text{tot}}(\text{PE)}$, and may therefore affect the \ArThirtyNine\ activity measurement.
The time constant of the triplet emission (we use \(\tau_{\text{slow}}\) for the measured value) was measured by performing an exponential fit of the late part of the waveform, i.e., between 1.1~$\mu$s and 5.7~$\mu$s after the trigger (see Fig.~\ref{fig:wfsamp}). Only ER-type events from the data are considered in this fit, i.e., events with $f_{\text{prompt}}<0.5$ and 200~PE$<Q_{\text{tot}}< 700$~PE.

\begin{figure}[h]
	\centering
	\begin{subfigure}{\columnwidth}
		\centering		\includegraphics[width=\columnwidth]{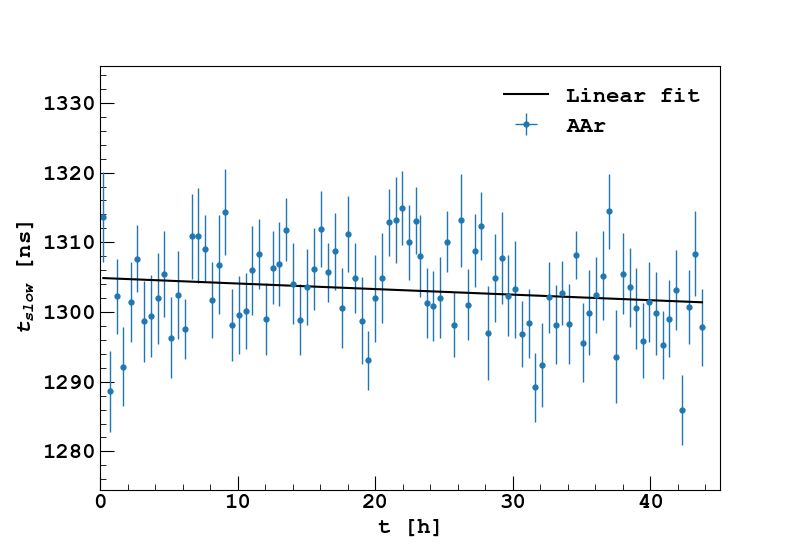}
		\caption{AAr}
	\end{subfigure}
	\begin{subfigure}{\columnwidth}
		\centering    \includegraphics[width=\columnwidth]{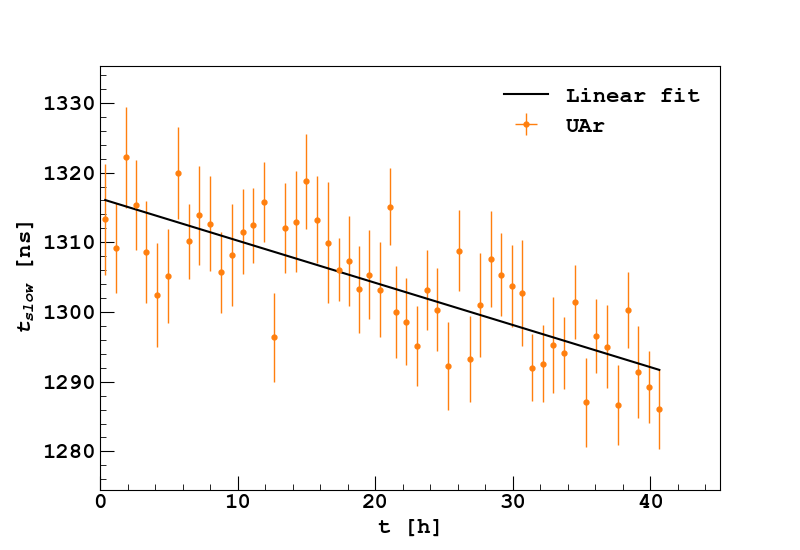}
		\caption{UAr}
	\end{subfigure}
	\caption{Measured time constant of the triplet emission, \(\tau_{\text{slow}}\)~(ns), vs time~(h) for (a) AAr; (b) UAr, with linear fits overlaid.}
	\label{fig:tripfinal}
\end{figure}

Fig.~\ref{fig:tripfinal} shows the \(\tau_{\text{slow}}\) value as a function of time for the AAr and UAr runs, respectively. 

The  corresponding live-times are  $t^{\text{live}}_\text{AAr}=(44.50\pm 0.01)$~h   and $t^{\text{live}}_\text{UAr}=(43.00\pm 0.01)$~h. With the average trigger rates, $5.94$~Hz in AAr and
$3.23$~Hz in UAr, the corresponding dead-time fractions are only $0.006\%$
and $0.003\%$, respectively.

The fitted linear slopes are ($-0.08\pm 0.05$)~ns/h and ($-0.60\pm 0.07$)~ns/h in the AAr and UAr runs, respectively, corresponding to a total variation of $\tau_\text{slow}$ over the run of 3.6~ns (0.27\%) and 25.8~ns (2.0\%), respectively. 

These variations are small and indicate that any increase in nitrogen contamination remains well below the ppm level~\cite{WArP:2008rgv}, and therefore adequate for our measurement. 


In both AAr and UAr runs, we observe no degradation of the integrated charge over time. 

The range of \(\tau_{\mathrm{slow}}\) values measured during the AAr run lies within that observed during the UAr run. Together with the stability of the SiPM gain, this supports consistent detector response between the two runs.



\section{Measurement of the \ArThirtyNine\  specific activity}
\label{sec:activity}

\begin{figure}[ht!]
	\centering
		\includegraphics[width=1\columnwidth]{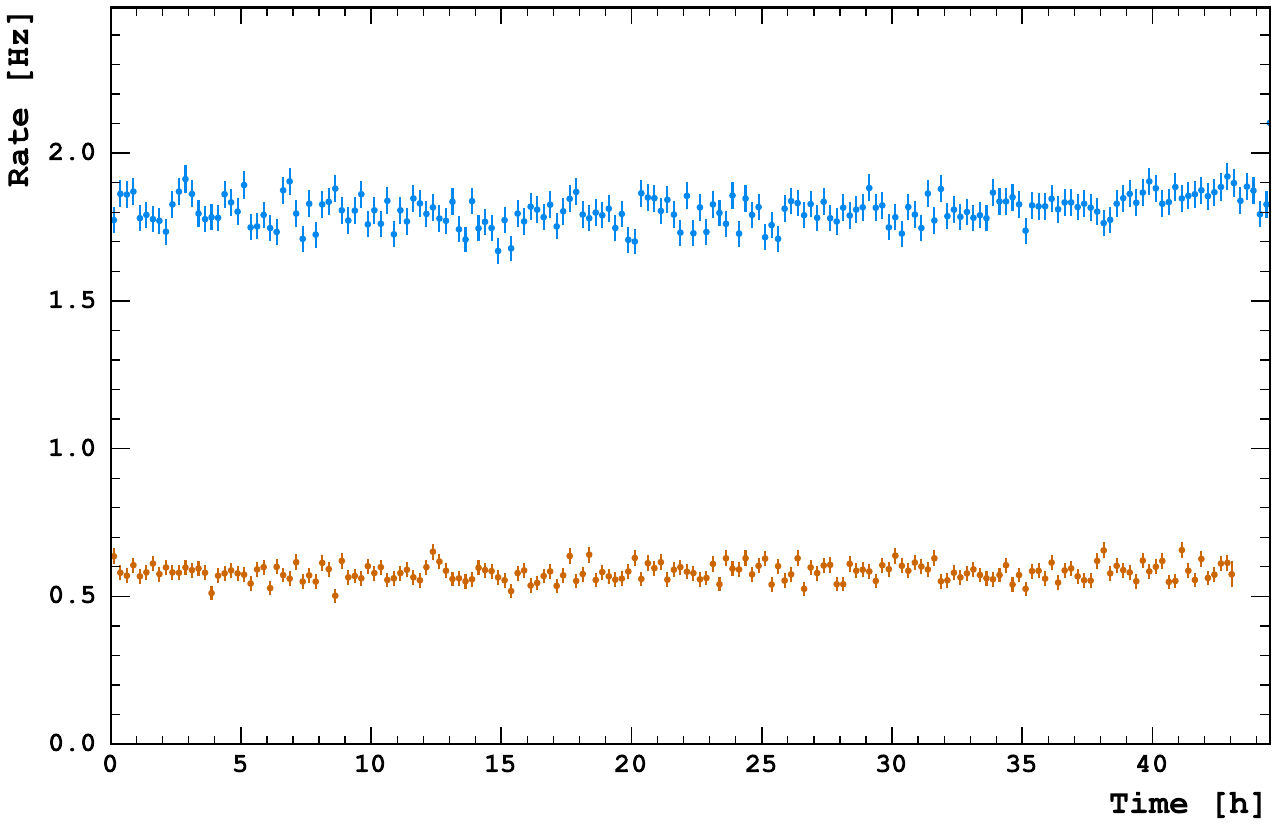}
        
	\caption{Rate of selected events for AAr (blue dots) and UAr (orange dots) vs. time, for $Q_{\text{tot}}(\text{PE)}$ below the upper analysis boundary of  250~PE. }
	\label{fig:evrate}
\end{figure}

Event selection rejects nuclear-recoil backgrounds, suppresses low-charge-noise events, and restricts the analysis to the reconstructed energy region below the \ce{^39Ar} endpoint.
Tab.~\ref{tab:final_cuts} lists the selection cuts used for the specific activity measurement. 
\begin{table}[ht!]
	\centering
    	\caption{Selection cuts for the specific activity measurement. The last cut is applied only to the cut-and-count method.}
	\begin{tabular}{lll}
		\hline
		cut      & value & units \\
        \hline 
        $f_\text{prompt}$ vs $Q_{\text{tot}}(\text{PE)}$ & as in Fig.~\ref{fig:PSD_AAr} & - \\
        Max($A$) & $>$5 per channel & mV \\
        $S_\text{f}$ \(Q^\text{T}_\text{tot}\) & $>$~2.5 & PE \\    
        $S_\text{f}$ $f_B$  \(Q^\text{B}_\text{tot}\) & $>$~2.5 & PE \\ 
        $Q_{\text{tot}}$ & $<$250 & PE  \\
        \hline
	\end{tabular}
	\label{tab:final_cuts}
\end{table}
Fig.~\ref{fig:evrate} shows the rate of selected events after cuts for AAr (blue dots) and UAr (orange dots) vs. time, for $Q_{\text{tot}}(\text{PE)}$ below the upper analysis boundary of  250~PE, and it is found to be stable. 



Since the goal of this study is the measurement of the \ArThirtyNine\  activity, our baseline method is a counting method, discussed in Sect~\ref{sec:counting}, to minimize the impact of the uncertainty from the theoretical spectrum and from the linearity of the $Q_{\text{tot}}(\text{PE)}$ response.
Our result will be compared with that obtained from
a maximum likelihood fit of the $Q_{\text{tot}}(\text{PE)}$ spectrum in Sect ~\ref{sec:like}.

 \subsection{Counting method}
 \label{sec:counting}



The specific activity of \(^{39}\)Ar in atmospheric argon is derived as:
\begin{equation}
{a}_{\text{AAr}} = \frac{N_{\text{AAr}}-N_{\text{UAr}} f_\text{AAr/UAr}}{\varepsilon_{\text{AAr}} \,t^\text{live}_\text{AAr} \, V_\text{LAr}\,\rho_\text{LAr}}+{a}^{^{39}\text{Ar}}_{\text{UAr}}+{a}^{^{85}\text{Kr}}_{\text{UAr}}+\mathrm{\Delta}{a}_{\text{AAr}}
,
\label{eq:spec_activity}
\end{equation}

 where $N_{\text{AAr}}$ is the number of selected events after cuts in the AAr run, $N_{\text{UAr}}$ is the number of selected events after cuts in the UAr run,  $f_\text{AAr/UAr}=t^\text{live}_\text{AAr}/t^\text{live}_\text{UAr}$ is the ratio of the AAr and UAr live-times, $\varepsilon_{\text{AAr}}$ is the total \ArThirtyNine\  event selection efficiency from simulation, $t^\text{live}_\text{AAr}$ is the atmospheric argon run duration, $V_\text{LAr}$ the liquid argon active volume, and $\rho_\text{LAr}$ the liquid argon density.  
The specific activities ${a}^{^{39}\text{Ar}}_{\text{UAr}}$ and ${a}^{^{85}\text{Kr}}_{\text{UAr}}$ are those reported in Eq.~\ref{eq:uar}.
 The isotope \ce{^85Kr} has a $\beta$-decay endpoint of 687~keV. Only approximately 2.5\% of the \(^{85}\mathrm{Kr}\) beta spectrum lies above the \(^{39}\mathrm{Ar}\) endpoint. Given the small \(^{85}\mathrm{Kr}\) activity in UAr, the contribution of this spectral tail is negligible for the present measurement. 
 
 The measured AAr activity depends on the AAr argon bottle's history. Argon activation by cosmic rays, which leads to equilibrium specific activity, occurs mainly in the upper atmosphere, and the equilibrium activity at sea level was estimated to be 800 times smaller~\cite{PhysRevC.100.024608}. It can be estimated that the \ce{^39Ar}  activity decreases by $\sim$2.1~mBq/kg per year.
 We reconstructed the history of our AAr bottle together with the producer. The argon bottle was filled on April 8$^{\text{th}}$, 2024, and gas distillation from air occurred $(30\pm 15)$~d before.
  The data run were taken on  May 7$^{\text{th}}$ and 8$^{\text{th}}$, 2024.
 Therefore, we added $\mathrm{\Delta} {a}_{\text{AAr}}=(0.34\pm 0.17)$~mBq/kg to take into account that 
  our AAr was distilled from air ($2\pm 1$)~m before the run. 

The active region is cylindrical, with a room temperature volume $V_\text{LAr}^\text{warm}$. 
We measured its dimensions using a Mitutoyo Absolute Coolant Proof IP67 CD-P20P caliper. The manufacturer-specified maximum permissible error is $\pm$40~$\mu$m for inner-diameter measurements and $\pm$20~$\mu$m for length measurements. Using NIST guidance to estimate instrument uncertainty, assuming a uniform distribution within $\pm a$, we obtain $a/\sqrt{3}$; for the length, this is 11.5~$\mu$m, and for the inner diameter, 23.1~$\mu$m, which we rounded to 20~$\mu$m overall.
We took nine measurements of the inner diameter and nine of the length.  The single-measurement precision (sample standard deviation) was 135~$\mu$m for length and 79~$\mu$m for inner diameter.
The uncertainty on the volume reported in Tab.~\ref{tab:acti_input} is obtained by propagating the standard errors of the mean dimensions, with the manufacturer-specified caliper accuracy added in quadrature.

During the AAr and UAr runs, the DArT volume was completely filled with liquid argon and so was the active volume.

The volume at liquid argon temperature  is given by:
\begin{equation}
    V_\text{LAr}=V_\text{LAr}^\text{warm}
    (1 + f_\text{cont})
\end{equation}
where $f_\text{cont}=\mathrm{\Delta}  {V}/{V}$ is the  fractional volume contraction. 
We measured the fractional linear contraction of acrylic using a leftover piece from a vessel machining run: a hollow cylinder 17.65~cm long, 9.65~cm in inner diameter, and 0.28~cm thick, measured at room temperature and liquid-nitrogen temperature. We performed ten measurements at room temperature and ten at liquid-nitrogen temperature for both length and inner diameter. The single-measurement precision from the statistical analysis ranged from 60~$\mu$m to 70~$\mu$m.
After immersion in liquid nitrogen, the cylinder was removed from the bath and
the cold dimensions were measured within a few seconds, so that no
significant thermal re-equilibration with the ambient environment was expected
during the measurement. The single-measurement precision from the statistical analysis ranged from 100~$\mu$m to 150~$\mu$m.
 We obtained $\mathrm{\Delta} {L}/{L}_\text{warm}=-(0.86\pm 0.07)\%$, where $L$ is the linear dimension, in agreement with data from Ref.~\cite{Hartwig1994}.
The correction from liquid argon to liquid nitrogen temperature, from the same reference, is approximately 0.02\%, which we neglect.
Assuming that our vessel is a mechanically unconstrained compact solid 3D  structure, since for small deformations, $
    \mathrm{\Delta} {V}/{V} = 3\mathrm{\Delta} {L}/{L}$,
we obtain $f_\text{cont}=-0.0258\pm 0.0021$. 

 The operating temperature for the two runs was 85~K with an uncertainty of $\pm$~1~K, limited by the sensor display's digit precision. The LAr density at 85~K, $\rho_\text{LAr}$, was taken from Ref.~\cite{10.1063/1.556037}, with an uncertainty of 0.2~\%. This value includes measurements from groups 1 and 2 as defined in the same paper. The density variation due to an uncertainty of $\pm$~1~K in absolute temperature was taken from Ref.~\cite{10.1063/1.555829} and is 0.4\%.
 The same temperature uncertainty has a negligible effect on the acrylic's thermal expansion coefficient.

The input factors for calculating the specific \ArThirtyNine\ activity, as given in Eq.~\ref{eq:spec_activity}, are listed in Tab.~\ref{tab:acti_input}. The efficiency uncertainty is the quadrature sum of the simulation's statistical uncertainty (0.023\%) and the systematic uncertainty due to the different $A^{\mathrm{corr}}_{\mathrm{TB}}$ distribution in data and simulation (0.14\%).

\begin{table}[ht!]
	
    \caption{Input factors to the calculation of the specific \ArThirtyNine\ activity of Eq.~\ref{eq:spec_activity}. First column: parameter.
Second column: units.
Third column:   parameter value.
Fourth column: relative uncertainty in \%. 
Fifth column: propagated relative uncertainty on the measured atmospheric-argon activity $a_{\mathrm{AAr}}$.
 Contributions to the uncertainty in ${a}_{\text{AAr}}$ that are smaller than 0.03\% are set to zero.}
 \centering
    \footnotesize
	\begin{tabular}{lcccc}
		\hline
        & units  & value  & unc. & unc. $\text{a}_{\text{AAr}}$\\
        & & & (\%) & (\%) \\
          \hline 
		  $N_{\text{AAr}}$    &  & 289662 &   & \\      
        $N_{\text{UAr}}$ & &94721  &   &  \\ \hline
        tot stat & & & & 0.33\\ \hline
        $f_\text{AAr/UAr}$ & & 1.035   & - & - \\ 
        $\varepsilon_{\text{AAr}}$& & 0.947 &  0.14 & 0.14 \\
        $t^\text{live}_\text{AAr}$ & s & 160200  &  - & - \\
 $\rho_\text{LAr}$ &  g/cm$^3$ & 1.403  & 0.44 & 0.44\\ 
 $V_\text{LAr}^\text{warm}$ &  cm$^3$ & 970.42  & 0.08 & 0.08 \\ 
 $f_\text{cont}$ &   & -0.0258  & 8 & 0.21\\ [3pt]
${a}^{^{85}\text{Kr}}_{\text{UAr}} $  
& mBq/kg &1.15  & 6 & -
     \\   [3pt]
${a}^{^{39}\text{Ar}}_{\text{UAr}}$ & mBq/kg &0.71  & 15 & - \\
     
 
     $\mathrm{\Delta}{a}_{\text{AAr}}$ & mBq/kg & 0.34 & 50 & - \\
     
     \hline
     tot syst (a) & & & & 0.51 \\ \hline
	\end{tabular}
	\label{tab:acti_input}
\end{table}

\begin{figure}[ht!]
\centering
\includegraphics[width=\columnwidth]{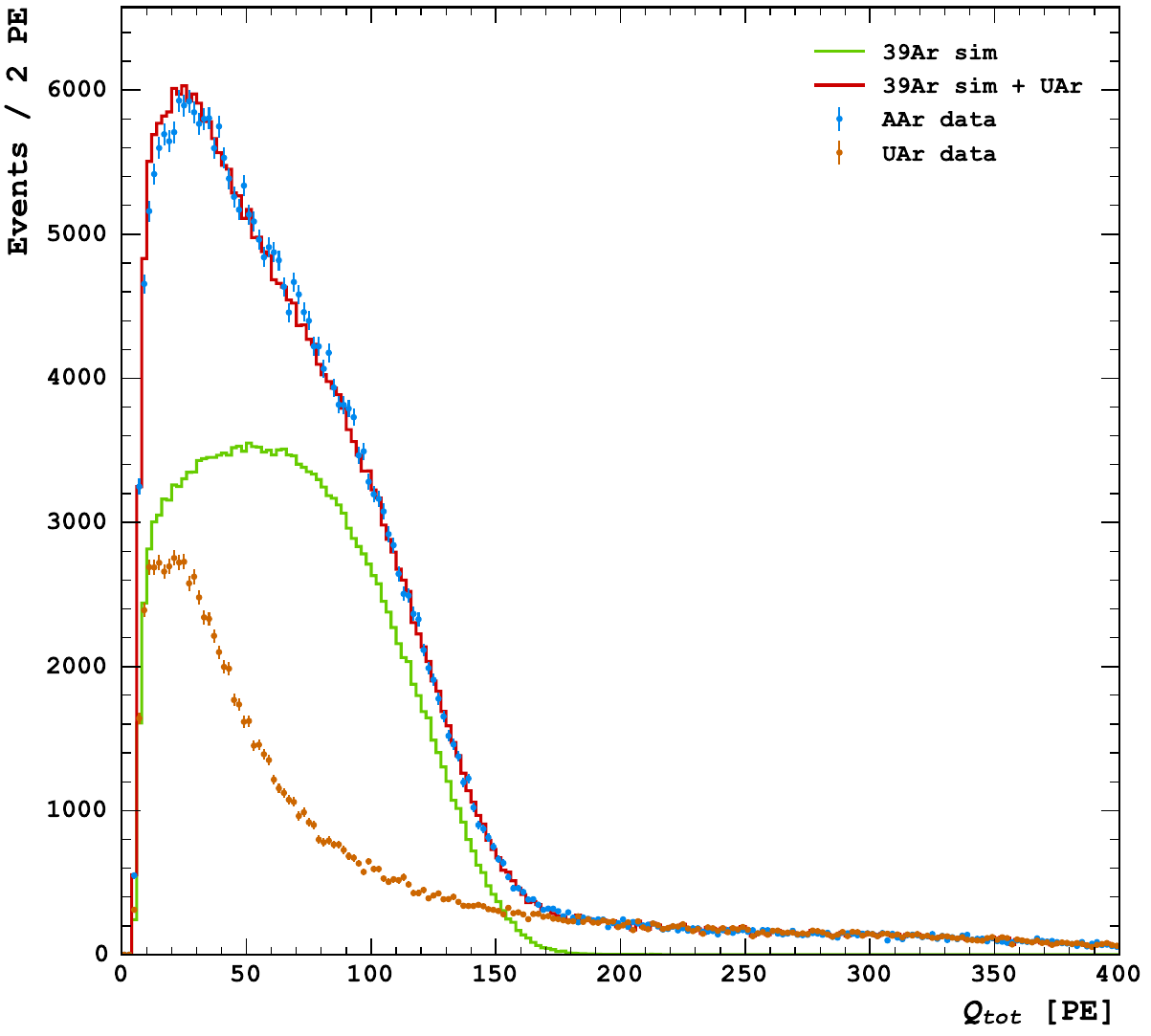}
\caption{\(Q_\text{{tot}}(\text{PE})\)  spectra for AAr data (blue points), UAr data
scaled to the AAr live-time (orange points), simulated
${}^{39}\mathrm{Ar}$ events (green solid line), and the sum
of the simulation and the scaled UAr spectrum (red solid line).
The simulated signal is normalized to the specific activity
reported in 
%
Eq.~\ref{eq:act_fin}.}
\label{fig:final_subtracted}
\centering
\end{figure}

Fig.~\ref{fig:final_subtracted} shows the \(Q_\text{{tot}}(\text{PE})\) spectra for AAr data, UAr data
scaled to the AAr live-time,  simulated
${}^{39}\mathrm{Ar}$ events, and the sum
of the simulation and the scaled UAr spectrum. The number of simulated events in the plot corresponds to the measured \ArThirtyNine\ specific activity, ${a}_{\text{AAr}}$, given by Eq.~\ref{eq:act_fin}.
In the \ArThirtyNine\ region, i.e., for \(Q_\text{{tot}}(\text{PE})\)  below 250~PE, neither of the two readout channels saturates.

We took the AAr and UAr runs under identical hardware conditions. In particular, we did not modify the detector or the data acquisition between the two fills, which occurred within a few days of each other. 

We tested the consistency of the non-\(^{39}\mathrm{Ar}\) backgrounds between the two runs by comparing the \(Q_\text{{tot}}(\text{PE})\) distributions after cuts above the \ce{^39Ar} $\beta$-decay endpoint, normalized by live-time. We selected events in the range $250-750~\mathrm{PE}$. The difference in the total number of events is  $\mathrm{\Delta}_\text{high}=58\pm 158$. When binning the distributions into 10~PE bins, the comparison between AAr and the scaled UAr spectrum gives $
\chi^2/n_\mathrm{dof} = 45.47/50 = 0.909$, with a pull ($P$) distribution characterized by
a mean of -0.054 and an RMS of  0.952, with 
$\max |P|$= 2.018, and 
only one out of 50 bins with 
$|P|>2$.

Tab.~\ref{tab:acti_syst} lists additional systematic uncertainties in the activity measurement. 
\begin{table}[ht!]
	\centering
    	\caption{Other systematic uncertainties in the activity measurement. First column: source or parameter associated with the systematic uncertainty.
Second column: nominal value of the parameter.
Third column: range over which the parameter is varied to evaluate the systematic effect.
Fourth column: propagated relative uncertainty on the measured atmospheric-argon activity $a_{\mathrm{AAr}}$. 
        Contributions smaller than 0.03\% are set to zero. The last line describes the quadratic sum of the systematic uncertainties of Tab.~\ref{tab:acti_input} and of this table.}
\footnotesize
	\begin{tabular}{lccc}
		\hline
		 & par. value & par. var. & unc. $a_{\mathrm{AAr}}$  \\
        & & &  (\%) \\
        \hline 
         asymm. factor data $f_\text{B}$  & \DArTCorrAsymFactDataMean & $\pm$ \DArTCorrAsymFactDataError & -- \\
         asymm. factor sim $f_\text{B}$  & \DArTCorrAsymFactSimMean & $\pm$ \DArTCorrAsymFactSimError & -- \\
         calibration factor $S_\text{f}$ & \DArTOverallCalibrationFactorMean & $\pm$ \DArTOverallCalibrationFactorError & -- \\
         thresh. cut in mV & $> 5~\mathrm{mV}$ & $\pm 2.5~\mathrm{mV}$ & -- \\
         thresh. cut in $Q^\text{T,B}_\text{tot}$ & $> 2.5~\mathrm{PE}$ & $\pm 0.5~\mathrm{PE}$ & 0.12 \\
         cut in max $Q_\text{tot}$ & $< 250~\mathrm{PE}$ & $\pm 20~\mathrm{PE}$ & 0.06 \\
         after-pulse, cross-talk & 10\%, 10\% & $\pm 10\%$ & -- \\
         SiPMs QE & 45\% & $\pm 5\%$ & 0.25 \\
         ESR reflectivity & 99\% & $\pm 0.5\%$ & 0.36 \\
         nuclear recoil backg. &  & & -- \\
         rise for $A^{\mathrm{corr}}_{\mathrm{TB}}>0.85$ &  &  & -- \\       
        $\beta$-spectrum shape &  &  & -- \\
        radon daughters &  &  & -- \\
         \hline 
         tot syst (b) & & & 0.47 \\
         tot syst ((a)$\oplus$(b)) & & & 0.69 \\
         \hline
	\end{tabular}
	\label{tab:acti_syst}
\end{table}

The asymmetry factors $f_\text{B}$ and the calibration factor $S_\text{f}$ were varied within their uncertainties; the lower cut values for amplitude and charge, and the upper cut value for charge, were also varied as described in Tab.~\ref{tab:acti_syst}.

Events due to electronic noise alone are suppressed by the
threshold cuts in Tab.~\ref{tab:final_cuts}. Any residual contribution is expected
to cancel in the live-time-normalized subtraction of the UAr
data from the AAr data, provided that the noise rate and
properties remain stable between the two runs. 
The cancellation may not be exact if one run exhibits higher noise at a given time. To verify this, we performed a noise-enhanced selection by requiring that one of the two channels have a signal amplitude below 5~mV. Still, the AAr and UAr runs exhibit a stable rate over time once the analysis cuts are applied. 

The effect of
electronic noise superimposed on genuine scintillation signals
is included in the waveform simulation and therefore accounted
for in the signal selection efficiency.

To evaluate the systematic uncertainty on the signal efficiency, we varied dark count rate, after-pulses, and cross-talk within their respective ranges in the simulation.

We also recomputed the efficiency by varying the reflectivity and quantum efficiency of the SiPMs independently over their respective ranges, and each time calculated a new $S_\text{f}$ to match the \(Q_\text{tot}(\text{PE})\) scale.


We tested the cancellation of nuclear-recoil backgrounds by comparing the live-time-normalized event counts above the selection boundary in Fig.~\ref{fig:PSD_AAr}, for \(Q_{\mathrm{tot}}<250\ \mathrm{PE}\). The difference was \(12\pm30\) events, consistent with zero.

We derive a relative systematic uncertainty in the specific activity by dividing the uncertainty of the above number by $N_{\text{AAr}}-N_{\text{UAr}} f_\text{AAr/UAr}$.


In the same way, we derive the relative uncertainty arising from the rise in the $A^{\mathrm{corr}}_{\mathrm{TB}}$ distribution above 0.85 (see Fig.~\ref{fig:TBA_comp}). The difference in the number of entries between AAr and UAr is $-20\pm 24$, compatible with zero.
 We derive a relative systematic uncertainty in the specific activity by dividing the uncertainty of the above number by $N_{\text{AAr}}-N_{\text{UAr}} f_\text{AAr/UAr}$.


The uncertainty arising from possible differences in the shape of the \ArThirtyNine\ decay spectrum in simulations was estimated by varying the theoretical formula of Ref.~\cite{MOUGEOT2023111018}  
within the authors' uncertainties, and found to be negligible.


Radon-related backgrounds can contribute to events in liquid-argon detectors through both $\alpha$- and $\beta$-decays~\cite{DarkSide:2016ddo, DEAP:2024mov}. In particular, the isotope \ce{^222Rn} 
can emanate from detector materials into the argon volume, and its decay products may then be present either in the liquid argon or plated out on detector surfaces~\cite{AMAUDRUZ2015178}. 

The presence of radon could contribute a systematic error in the \ce{^39Ar}  specific activity measurement if its content in the AAr and UAr were different, and therefore lead to a different number of $\beta$-decays in the \ce{^39Ar} decay region. The prompt part of the \ce{^222Rn} chain gives rise to three $\alpha$-decays, i.e., $^{222}$Rn, $^{218}$Po,  $^{214}$Po, and two $\beta$-decays,  \ce{^214Pb}  and $^{214}$Bi, in secular equilibrium.
We measured the total $\alpha$-particle rate and derived the $\beta$-decay rate by scaling it by 2/3, assuming full detection efficiency for the $\alpha$-decay, and by the fraction of the $\beta$-decay spectra that lies below 250~PE.

\begin{figure}[ht!]
\centering
\includegraphics[width=\columnwidth]{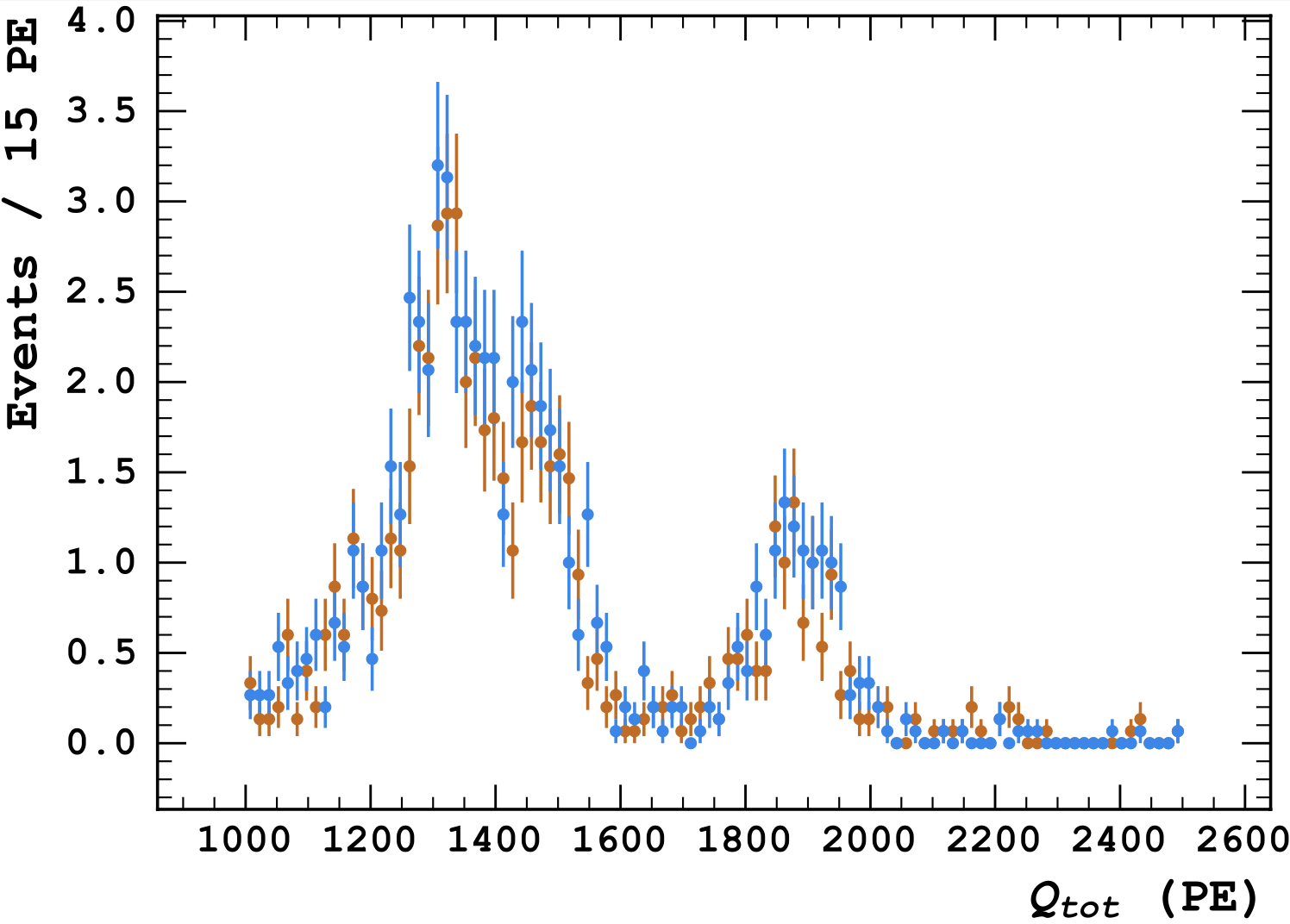}
\caption{$Q_{\mathrm{tot}}$(PE) spectrum for $Q_{\mathrm{tot}} > 1000$~PE in the AAr (blue dots) and UAr (orange dots) data. 
}
\label{fig:dart_bipo_UAr}
\end{figure}

In Fig.~\ref{fig:dart_bipo_UAr}, we show the $Q_{\mathrm{tot}}$(PE)  spectrum for  AAr and UAr, for $Q_{\mathrm{tot}} > 1000$~PE . Due to saturation in the $Q_{\mathrm{tot}}$(PE) response at high values, the f$_{\text{prompt}}$ cut was not applied to this selection. 
The total number of events is $1017 \pm 32$ in AAr and $921 \pm 30$ in UAr.

The expected number of $\beta$-events in the $^{39}$Ar energy region, after accounting for the different endpoint energies and the detection efficiency, is $520 \pm 16$ in AAr and $470 \pm 15$ in UAr. 
The difference in $\beta$-decays between AAr and UAr, rescaled by the live-time,  is therefore $34\pm 23$ events. 
We derive a relative systematic uncertainty in the specific activity by dividing this difference by $N_{\text{AAr}}-N_{\text{UAr}}f_\text{AAr/UAr}$, and it turns out to be negligible. 

From the comparison of the peak positions in the UAr and  AAr distributions, we also establish that the absolute $Q_{\mathrm{tot}}$(PE) scale agrees in the two runs to better than 1\%.

The  result is: 
\begin{equation}
{a}_{\text{AAr}} = 0.955 \pm 0.003_{\text{stat.}} \pm 0.007_{\text{syst.}}~\text{Bq}/\text{kg}. 
\label{eq:act_fin}
\end{equation}

By summing the statistical and the systematic error in quadrature,  
\begin{equation}
{a}_{\text{AAr}} = 0.955 \pm 0.008_{\text{tot.}}~\text{Bq}/\text{kg}. 
\label{eq:act_fin_spec}
\end{equation}

\subsection{Likelihood fit}
\label{sec:like}
The measurement of the \ce{^39}Ar specific activity was also performed with an extended binned maximum-likelihood fit using the \textsc{RooFit} framework~\cite{verkerke2003roofittoolkitdatamodeling}. The likelihood fit is performed in the range
$5~\mathrm{PE}<Q_{\mathrm{tot}}<300~\mathrm{PE}$ using 60 equal-width bins.
In each bin, the observed AAr count is
described by a Poisson term whose expectation is the sum of a UAr-driven
background component and a simulated \ce{^39Ar} signal component. The UAr
spectrum is therefore not treated as an exact fixed template: its finite
bin-by-bin statistical uncertainty is propagated through constrained nuisance
parameters. 
We considered two kinds of systematic uncertainties separately. The uncertainties that change only the normalization of the \ce{^39}Ar spectrum in the simulation, those of Tab.~\ref{tab:acti_input}, were propagated into the fit as Gaussian nuisance parameters. 

The \(^{39}\mathrm{Ar}\) and \(^{85}\mathrm{Kr}\) activities in UAr were constrained by Gaussian terms centered on the values reported in Sect.~\ref{sec:Uar}, with widths equal to their quoted uncertainties.


The finite Monte Carlo statistics of the signal template were included through bin-by-bin Gaussian nuisance parameters. The quoted correlations come from the fit covariance matrix.
For the Gaussian-constrained nuisance parameters, the pull is defined as the fitted displacement divided by its prior uncertainty. 
The nominal numbers of degrees of freedom are 58 and 56 for the two-parameter and four-parameter fits, respectively, calculated as the number of bins minus the number of global fit parameters.
The fit quality is evaluated from the post-fit residuals between the observed AAr counts, $n_i$, and the fitted expectation, $\widehat{\mu}_i$. The pull in each bin is defined as
\begin{equation}
p_i=
\frac{n_i-\widehat{\mu}_i}
{\sqrt{\widehat{\mu}_i}},
\label{eq:pull}
\end{equation}
where the denominator is the Poisson standard deviation evaluated at the fitted expectation. The reported Pearson chi-square is obtained by summing the squared bin pulls,
\begin{equation}
\chi^2_{\mathrm{P}}=
\sum_i p_i^2
=
\sum_i
\frac{(n_i-\widehat{\mu}_i)^2}
{\widehat{\mu}_i}.
\end{equation}

We obtained:
\begin{equation}
{a}_{\text{AAr}} =  0.958 \pm 0.005_{\text{fit}}~\text{Bq}/\text{kg}  
\label{eq:act_fin_fit}
\end{equation} 
with $S_\text{f}  =  1.034\pm 0.002$.  

\begin{figure}[ht!]
\centering
\includegraphics[width=\columnwidth]{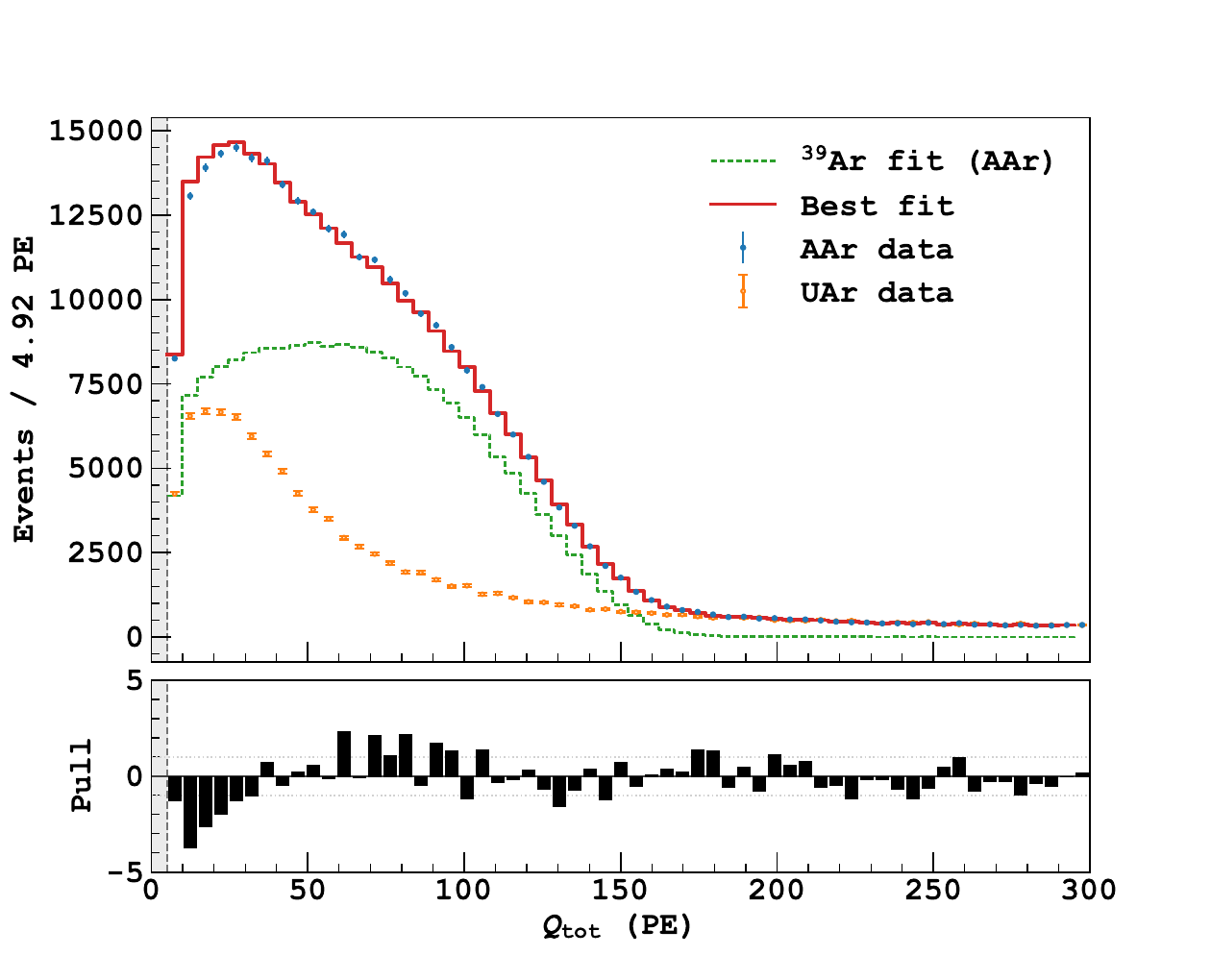}
\caption{\(Q_\text{tot}(\text{PE})\) spectrum 
for AAr data (blue points) and UAr data
scaled to the AAr live time (orange points).
The red solid line shows the best-fit total prediction from
the fit with two global parameters, and the green dashed line
shows its ${}^{39}\mathrm{Ar}$ signal component.
The lower panel shows the bin pulls, as defined in Eq.~\ref{eq:pull}.}
\label{fig:act_fin_fit}
\centering
\end{figure}

The Pearson $\chi^2/n_\text{dof}$ is 85.5/58~=~1.47 and the correlation coefficient  $r_{{a}_{\text{AAr}},S_f}=-0.007$. The fit result is shown in Fig.~\ref{fig:act_fin_fit}.

The dominant contribution to $\chi^2/n_\text{dof}$ arises from the low-energy region of the spectrum. 
Deviations between the measured and simulated \(^{39}\)Ar spectra cannot be directly interpreted as deficiencies of the underlying $\beta$-decay calculation. The measured distribution results from convolving the physical $\beta$-spectrum with the detector energy response, resolution, and selection efficiency. 

To further investigate the consequences of this discrepancy on the activity
measurement, we introduce an energy-dependent phenomenological correction to
the simulated \ce{^{39}Ar} spectrum. A linear correction was insufficient to reproduce the low-charge region.
The charge in the  simulation is therefore parametrized as
\begin{equation}
Q_{\mathrm{tot}} =
Q^0_{\mathrm{tot}}
\left(S_\text{f} + b\,u + c\,u^2\right),
\label{eq:four_fit}
\end{equation}
where
\begin{equation}
u =
\frac{
Q^0_{\mathrm{tot}} - Q^{0,\mathrm{mid}}_{\mathrm{tot}}
}{
Q^{0,\max}_{\mathrm{tot}} - Q^{0,\min}_{\mathrm{tot}}
}.
\label{eq:four_fit2}
\end{equation}
Here, \(Q^{0,\mathrm{mid}}_{\mathrm{tot}}\) denotes the center of the fitted
interval, while \(Q^{0,\min}_{\mathrm{tot}}\) and \(Q^{0,\max}_{\mathrm{tot}}\)
are the lower and upper fit boundaries. The variable \(u\) is therefore
dimensionless and centered in the fitted range. The parameters \(b\) and \(c\)
describe a phenomenological first- and second-order deformation of the charge
response.

\begin{figure}[ht!]
\centering
\includegraphics[width=\columnwidth]{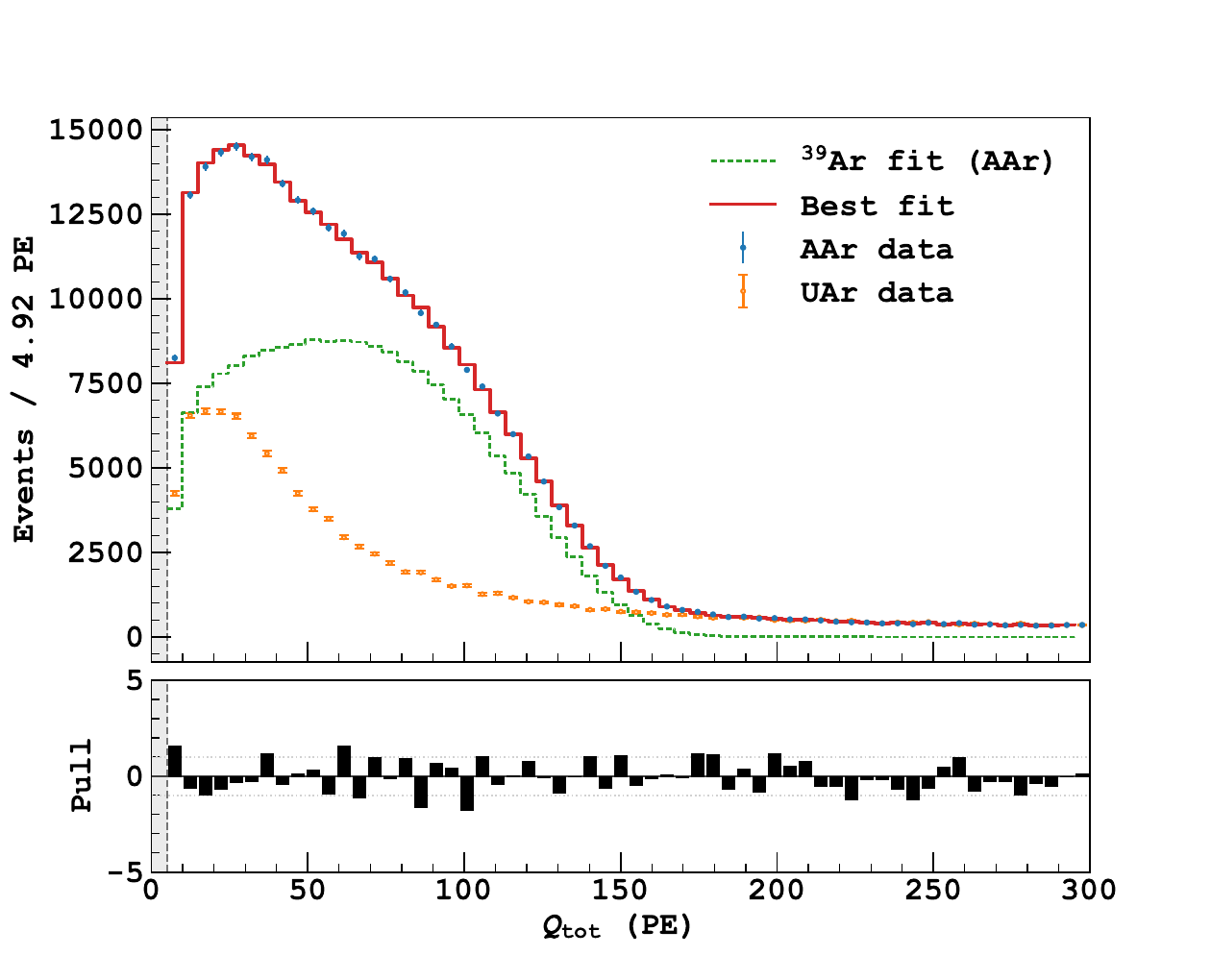}
\caption{\(Q_\text{tot}(\text{PE})\) spectrum, as defined in Eq.~\ref{eq:four_fit} and Eq.~\ref{eq:four_fit2}, 
for AAr data (blue points) and UAr data
scaled to the AAr live time (orange points).
The red solid line shows the best-fit total prediction from
the fit with four global parameters, and the green dashed line
shows its ${}^{39}\mathrm{Ar}$ signal component.
 The lower panel shows the bin pulls, as defined in Eq.~\ref{eq:pull}.}
\label{fig:act_fin_fit2}
\centering
\end{figure}

The quadratic fit gives
\begin{equation}
{a}_{\text{AAr}} =  0.954 \pm 0.008_{\text{fit}}~\text{Bq}/\text{kg}  
\label{eq:act_fin_fit_quadratic}
\end{equation}
with $S_\text{f}  =  1.0027 \pm 0.0029$,
$b  =  0.0869 \pm 0.0157$, 
 and $c  =  0.395 \pm 0.035$, 
with $\chi^2/n_\text{dof} = 41.8/56 = 0.75$,
and correlation coefficients 
$r_{a_{\mathrm{AAr}},S_f} = 0.015$, 
$r_{a_{\mathrm{AAr}},b} = 0.055$, 
$r_{a_{\mathrm{AAr}},c} = -0.042$.
The fit result is shown in Fig.~\ref{fig:act_fin_fit2}.
The additional response parameters substantially improve the agreement with the data. The fitted activity decreases by 4~mBq/kg relative to the two-parameter fit and differs from the cut-and-count result by only 1~mBq/kg, while the fit uncertainty increases from 5 to \(8\ \mathrm{mBq/kg}\).

For uncertainties in factors that affect the \ce{^39}Ar spectrum shape, as discussed in Tab.~\ref{tab:acti_syst}, we repeated the fit after applying the corresponding upward and downward variation. We assigned the largest absolute shift relative to the nominal result as the systematic uncertainty. We assumed the individual contributions were independent and combined them in quadrature. We did not double-count uncertainties already included as nuisance parameters. The result is summarized in Tab.~\ref{tab:acti_syst_fit}.
These systematic uncertainties were finally combined in quadrature with the quoted fit uncertainty of Eq.~\ref{eq:act_fin_fit_quadratic}.
As expected, the likelihood method is much more sensitive to these uncertainties than the cut-and-count method.
\begin{table}[ht!]
	\centering
	\caption{Factors leading to systematic uncertainties on the activity measurement related to the \ce{^39Ar} spectrum shape changes with the likelihood method: parameter value, range of variation, and uncertainty on $a_{\mathrm{AAr}}$.}
	\renewcommand{\arraystretch}{1.15}
    \footnotesize
	\begin{tabular}{lccc}
		\hline
		type & par. value & par. var. & unc. $a_{\mathrm{AAr}}$ (\%) \\
		\hline
		SiPMs QE & 45\% & $\pm 5\%$ & 0.7 \\
		ESR reflectivity & 99\% & $\pm 0.5\%$ & 1.0 \\
		\hline
		tot syst (fit) & & & 1.3 \\
		\hline
	\end{tabular}
	\label{tab:acti_syst_fit}
\end{table}
Including all uncertainties, we obtain:
\begin{equation}
{a}_{\text{AAr}} =  0.954 \pm 0.015_{\text{tot}}~\text{Bq}/\text{kg}  
\end{equation}

\subsection{Discussion}

We take the cut-and-count result in Eq.~\ref{eq:act_fin_spec} as our final result, owing to its small dependence
on the \ce{^39}Ar spectrum shape, 

Our result is consistent with previous measurements of the \ce{^39Ar} specific activity, with our final uncertainty being about three times smaller than the current most precise measurement from the DEAP-3600 experiment.



\section*{Acknowledgments}
We acknowledge the continuous support of the Canfranc Underground Laboratory and of its personnel.
This work was supported by the Istituto Nazionale di Fisica Nucleare (INFN) under the Darkside-20k project. We acknowledge the support from FSC 2014-2020 - Patto per lo Sviluppo, Regione Sardegna, Italy.
This work was made possible by funding from the Spanish Ministry of Science and Innovation (MCIN) under Grant PID2022-138357NB-C22.

\section*{Data Availability Statement}
The data supporting the findings of this study are available
from the corresponding author upon request.

\section*{Code Availability Statement}

The code supporting the findings of this study is available
from the corresponding author upon request.

\section*{Open Access} 

This article is licensed under a Creative Commons Attribution 4.0 International License, which permits use, sharing, adaptation,
distribution and reproduction in any medium or format, as long as you
give appropriate credit to the original author(s) and the source, provide a link to the Creative Commons license, and indicate if changes
were made. The images or other third-party material in this article
are included in the article’s Creative Commons license, unless indicated otherwise in a credit line to the material. If material is not
included in the article’s Creative Commons license and your intended
use is not permitted by statutory regulation or exceeds the permitted use, you will need to obtain permission directly from the copyright holder. To view a copy of this license, visit \url{http://creativecommons.org/licenses/by/4.0/}.
Funded by SCOAP$^3$. SCOAP$^3$ supports the goals of the International
Year of Basic Sciences for Sustainable Development.

\bibliographystyle{ds}
\bibliography{wal, bibliography}

\begin{thebibliography}{40}
\expandafter\ifx\csname natexlab\endcsname\relax\def\natexlab#1{#1}\fi
\expandafter\ifx\csname bibnamefont\endcsname\relax
  \def\bibnamefont#1{#1}\fi
\expandafter\ifx\csname bibfnamefont\endcsname\relax
  \def\bibfnamefont#1{#1}\fi
\expandafter\ifx\csname citenamefont\endcsname\relax
  \def\citenamefont#1{#1}\fi
\expandafter\ifx\csname url\endcsname\relax
  \def\url#1{\texttt{#1}}\fi
\expandafter\ifx\csname urlprefix\endcsname\relax\def\urlprefix{URL }\fi
\providecommand{\bibinfo}[2]{#2}
\providecommand{\eprint}[2][]{\url{#2}}

\bibitem[{\citenamefont{Aalseth et~al.}(2018)}]{Aalseth:2018gq}
\bibinfo{author}{\bibfnamefont{C.~E.} \bibnamefont{Aalseth}}  \bibnamefont{et~al.},  \href{http://dx.doi.org/10.1140/epjp/i2018-11973-4}{\bibinfo{journal}{Eur. Phys. J. Plus} \textbf{\bibinfo{volume}{133}}, \bibinfo{pages}{131}\bibinfo{year}{ (\bibinfo{year}{2018})}}.

\bibitem[{\citenamefont{Abi et~al.}(2020)}]{DUNE:2020txw}
\bibinfo{author}{\bibfnamefont{B.}~\bibnamefont{Abi}} \bibnamefont{et~al.},  \href{http://dx.doi.org/10.1088/1748-0221/15/08/T08010}{\bibinfo{journal}{JINST} \textbf{\bibinfo{volume}{15}}, \bibinfo{pages}{T08010}\bibinfo{year}{ (\bibinfo{year}{2020})}}.

\bibitem[{\citenamefont{Abed~Abud et~al.}(2024)}]{DUNE:2024wvj}
\bibinfo{author}{\bibfnamefont{A.}~\bibnamefont{Abed~Abud}} \bibnamefont{et~al.},  \href{http://dx.doi.org/10.1088/1748-0221/19/12/P12005}{\bibinfo{journal}{JINST} \textbf{\bibinfo{volume}{19}}, \bibinfo{pages}{P12005}\bibinfo{year}{ (\bibinfo{year}{2024})}}.

\bibitem[{\citenamefont{Adhikari et~al.}(2025{\natexlab{a}})}]{DEAP:2025shk}
\bibinfo{author}{\bibfnamefont{P.}~\bibnamefont{Adhikari}} \bibnamefont{et~al.},  \href{http://dx.doi.org/10.1140/epjc/s10052-025-14289-5}{\bibinfo{journal}{Eur. Phys. J. C} \textbf{\bibinfo{volume}{85}}, \bibinfo{pages}{728}\bibinfo{year}{ (\bibinfo{year}{2025}{\natexlab{a}})}}.

\bibitem[{\citenamefont{Saldanha et~al.}(2019)}]{PhysRevC.100.024608}
\bibinfo{author}{\bibfnamefont{R.}~\bibnamefont{Saldanha}}  \bibnamefont{et~al.},  \href{http://dx.doi.org/10.1103/PhysRevC.100.024608}{\bibinfo{journal}{Phys. Rev. C} \textbf{\bibinfo{volume}{100}}, \bibinfo{pages}{024608}\bibinfo{year}{ (\bibinfo{year}{2019})}}.

\bibitem[{\citenamefont{Bhattacharya et~al.}(2026)}]{Bhattacharya:2025emx}
\bibinfo{author}{\bibfnamefont{S.}~\bibnamefont{Bhattacharya}} \bibnamefont{et~al.},  \href{http://dx.doi.org/10.1016/j.gca.2025.12.041}{\bibinfo{journal}{Geochim. Cosmochim. Acta} \textbf{\bibinfo{volume}{415}}, \bibinfo{pages}{196}\bibinfo{year}{ (\bibinfo{year}{2026})}}.

\bibitem[{\citenamefont{Loosli}(1983)}]{Loosli1983Ar39Dating}
\bibinfo{author}{\bibfnamefont{H.~H.} \bibnamefont{Loosli}},  \href{http://dx.doi.org/10.1016/0012-821X(83)90021-3}{\bibinfo{journal}{Earth Planet. Sci. Lett.} \textbf{\bibinfo{volume}{63}}, \bibinfo{pages}{51}\bibinfo{year}{ (\bibinfo{year}{1983})}}.

\bibitem[{\citenamefont{Ritterbusch et~al.}(2014)}]{Ritterbusch2014Ar39ATTA}
\bibinfo{author}{\bibfnamefont{F.}~\bibnamefont{Ritterbusch}}  \bibnamefont{et~al.},  \href{http://dx.doi.org/10.1002/2014GL061120}{\bibinfo{journal}{Geophys. Res. Lett.} \textbf{\bibinfo{volume}{41}}, \bibinfo{pages}{6758}\bibinfo{year}{ (\bibinfo{year}{2014})}}.

\bibitem[{\citenamefont{Adhikari et~al.}(2023)}]{DEAP:2023wri}
\bibinfo{author}{\bibfnamefont{P.}~\bibnamefont{Adhikari}} \bibnamefont{et~al.},  \href{http://dx.doi.org/10.1140/epjc/s10052-023-11678-6}{\bibinfo{journal}{Eur. Phys. J. C} \textbf{\bibinfo{volume}{83}}, \bibinfo{pages}{642}\bibinfo{year}{ (\bibinfo{year}{2023})}}.

\bibitem[{\citenamefont{Benetti et~al.}(2007)}]{WARP:2006nsa}
\bibinfo{author}{\bibfnamefont{P.}~\bibnamefont{Benetti}} \bibnamefont{et~al.},  \href{http://dx.doi.org/10.1016/j.nima.2007.01.106}{\bibinfo{journal}{Nucl. Instrum. Meth. A} \textbf{\bibinfo{volume}{574}}, \bibinfo{pages}{83}\bibinfo{year}{ (\bibinfo{year}{2007})}}.

\bibitem[{\citenamefont{Calvo et~al.}(2018)}]{ArDM:2017ndf}
\bibinfo{author}{\bibfnamefont{J.}~\bibnamefont{Calvo}} \bibnamefont{et~al.},  \href{http://dx.doi.org/10.1088/1475-7516/2018/12/011}{\bibinfo{journal}{JCAP} \textbf{\bibinfo{volume}{12}}, \bibinfo{pages}{011}\bibinfo{year}{ (\bibinfo{year}{2018})}}.

\bibitem[{\citenamefont{Aalseth et~al.}(2020)}]{DarkSide-20k:2020qfz}
\bibinfo{author}{\bibfnamefont{C.~E.} \bibnamefont{Aalseth}} \bibnamefont{et~al.},  \href{http://dx.doi.org/10.1088/1748-0221/15/02/P02024}{\bibinfo{journal}{JINST} \textbf{\bibinfo{volume}{15}}, \bibinfo{pages}{P02024}\bibinfo{year}{ (\bibinfo{year}{2020})}}.

\bibitem[{\citenamefont{Acerbi et~al.}(2025{\natexlab{a}})}]{DarkSide-20k:2024inx}
\bibinfo{author}{\bibfnamefont{F.}~\bibnamefont{Acerbi}} \bibnamefont{et~al.},  \href{http://dx.doi.org/10.1088/1748-0221/20/02/P02016}{\bibinfo{journal}{JINST} \textbf{\bibinfo{volume}{20}}, \bibinfo{pages}{P02016}\bibinfo{year}{ (\bibinfo{year}{2025}{\natexlab{a}})}}.

\bibitem[{\citenamefont{Agnes et~al.}(2024)}]{GlobalArgonDarkMatter:2024wtv}
\bibinfo{author}{\bibfnamefont{P.}~\bibnamefont{Agnes}} \bibnamefont{et~al.},  \href{http://dx.doi.org/10.3389/fphy.2024.1387069}{\bibinfo{journal}{Front. Phys.} \textbf{\bibinfo{volume}{12}}, \bibinfo{pages}{1387069}\bibinfo{year}{ (\bibinfo{year}{2024})}}.

\bibitem[{\citenamefont{Agnes et~al.}(2021)}]{DarkSide-20k:2021nia}
\bibinfo{author}{\bibfnamefont{P.}~\bibnamefont{Agnes}} \bibnamefont{et~al.},  \href{http://dx.doi.org/10.1140/epjc/s10052-021-09121-9}{\bibinfo{journal}{Eur. Phys. J. C} \textbf{\bibinfo{volume}{81}}, \bibinfo{pages}{359}\bibinfo{year}{ (\bibinfo{year}{2021})}}.

\bibitem[{\citenamefont{Aaron et~al.}(2023)}]{DarkSide-20k:2023grj}
\bibinfo{author}{\bibfnamefont{E.}~\bibnamefont{Aaron}} \bibnamefont{et~al.},  \href{http://dx.doi.org/10.1140/epjc/s10052-023-11430-0}{\bibinfo{journal}{Eur. Phys. J. C} \textbf{\bibinfo{volume}{83}}, \bibinfo{pages}{453}\bibinfo{year}{ (\bibinfo{year}{2023})}}.

\bibitem[{\citenamefont{Agnes et~al.}(2018{\natexlab{a}})}]{DarkSide:2018kuk}
\bibinfo{author}{\bibfnamefont{P.}~\bibnamefont{Agnes}} \bibnamefont{et~al.},  \href{http://dx.doi.org/10.1103/PhysRevD.98.102006}{\bibinfo{journal}{Phys. Rev. D} \textbf{\bibinfo{volume}{98}}, \bibinfo{pages}{102006}\bibinfo{year}{ (\bibinfo{year}{2018}{\natexlab{a}})}}.

\bibitem[{\citenamefont{Agnes et~al.}(2018{\natexlab{b}})}]{DarkSide:2018bpj}
\bibinfo{author}{\bibfnamefont{P.}~\bibnamefont{Agnes}} \bibnamefont{et~al.},  \href{http://dx.doi.org/10.1103/PhysRevLett.121.081307}{\bibinfo{journal}{Phys. Rev. Lett.} \textbf{\bibinfo{volume}{121}}, \bibinfo{pages}{081307}\bibinfo{year}{ (\bibinfo{year}{2018}{\natexlab{b}})}}.

\bibitem[{\citenamefont{Gola et~al.}(2019)}]{Gola2019}
\bibinfo{author}{\bibfnamefont{A.}~\bibnamefont{Gola}} \bibnamefont{et~al.},  \href{http://dx.doi.org/10.3390/s19020308}{\bibinfo{journal}{Sensors} \textbf{\bibinfo{volume}{19}}, \bibinfo{pages}{308}\bibinfo{year}{ (\bibinfo{year}{2019})}}.

\bibitem[{\citenamefont{Acerbi et~al.}(2025{\natexlab{b}})}]{DarkSide-20k:2025avf}
\bibinfo{author}{\bibfnamefont{F.}~\bibnamefont{Acerbi}} \bibnamefont{et~al.},  \href{http://dx.doi.org/10.1140/epjc/s10052-025-14940-1}{\bibinfo{journal}{Eur. Phys. J. C} \textbf{\bibinfo{volume}{85}}, \bibinfo{pages}{1334}\bibinfo{year}{ (\bibinfo{year}{2025}{\natexlab{b}})}}.

\bibitem[{\citenamefont{Organtini et~al.}(2020)}]{Organtini:2020bga}
\bibinfo{author}{\bibfnamefont{P.}~\bibnamefont{Organtini}}  \bibnamefont{et~al.},  \href{http://dx.doi.org/10.1016/j.nima.2020.164410}{\bibinfo{journal}{Nucl. Instrum. Meth. A} \textbf{\bibinfo{volume}{978}}, \bibinfo{pages}{164410}\bibinfo{year}{ (\bibinfo{year}{2020})}}.

\bibitem[{\citenamefont{D'Incecco et~al.}(2018)}]{DIncecco2:2018hy}
\bibinfo{author}{\bibfnamefont{M.}~\bibnamefont{D'Incecco}}  \bibnamefont{et~al.},  \href{http://dx.doi.org/10.1109/TNS.2017.2774779}{\bibinfo{journal}{IEEE Trans. Nucl. Sci.} \textbf{\bibinfo{volume}{65}}, \bibinfo{pages}{591}\bibinfo{year}{ (\bibinfo{year}{2018})}}.

\bibitem[{\citenamefont{Kugathasan}(2020)}]{Kugathasan:2020xry}
\bibinfo{author}{\bibfnamefont{R.}~\bibnamefont{Kugathasan}},  \href{http://dx.doi.org/10.22323/1.370.0011}{\bibinfo{journal}{PoS} \textbf{\bibinfo{volume}{TWEPP2019}}, \bibinfo{pages}{011}\bibinfo{year}{ (\bibinfo{year}{2020})}}.

\bibitem[{\citenamefont{Mougeot}(2023)}]{MOUGEOT2023111018}
\bibinfo{author}{\bibfnamefont{X.}~\bibnamefont{Mougeot}},  \href{http://dx.doi.org/10.1016/j.apradiso.2023.111018}{\bibinfo{journal}{Appl. Radiat. Isot.} \textbf{\bibinfo{volume}{201}}, \bibinfo{pages}{111018}\bibinfo{year}{ (\bibinfo{year}{2023})}}.

\bibitem[{Win(2018)}]{WinNT}
\emph{\bibinfo{title}{{\em 3MTM Enhanced Specular Reflector (ESR)} application guidelines}} (\bibinfo{year}{2018}), \urlprefix\url{https://multimedia.3m.com/mws/media/1389248O/application-guide-for-esr.pdf}.

\bibitem[{\citenamefont{Boulay et~al.}(2021)}]{Boulay:2021njr}
\bibinfo{author}{\bibfnamefont{M.~G.} \bibnamefont{Boulay}} \bibnamefont{et~al.},  \href{http://dx.doi.org/10.1140/epjc/s10052-021-09870-7}{\bibinfo{journal}{Eur. Phys. J. C} \textbf{\bibinfo{volume}{81}}, \bibinfo{pages}{1099}\bibinfo{year}{ (\bibinfo{year}{2021})}}.

\bibitem[{\citenamefont{Agnes et~al.}(2016)}]{DarkSide:2015cqb}
\bibinfo{author}{\bibfnamefont{P.}~\bibnamefont{Agnes}}  \bibnamefont{et~al.},  \href{http://dx.doi.org/10.1103/PhysRevD.93.081101}{\bibinfo{journal}{Phys. Rev. D} \textbf{\bibinfo{volume}{93}}, \bibinfo{pages}{081101(R)}\bibinfo{year}{ (\bibinfo{year}{2016})}}.

\bibitem[{\citenamefont{Lindemann and Simgen}(2014)}]{Lindemann:2013kna}
\bibinfo{author}{\bibfnamefont{S.}~\bibnamefont{Lindemann}} \bibnamefont{and} \bibinfo{author}{\bibfnamefont{H.}~\bibnamefont{Simgen}},  \href{http://dx.doi.org/10.1140/epjc/s10052-014-2746-1}{\bibinfo{journal}{Eur. Phys. J. C} \textbf{\bibinfo{volume}{74}}, \bibinfo{pages}{2746}\bibinfo{year}{ (\bibinfo{year}{2014})}}.

\bibitem[{\citenamefont{Otono et~al.}(2006)}]{Otono:2006zz}
\bibinfo{author}{\bibfnamefont{H.}~\bibnamefont{Otono}}  \bibnamefont{et~al.},  \href{http://dx.doi.org/10.22323/1.051.0007}{\bibinfo{journal}{PoS} \textbf{\bibinfo{volume}{PD07}}, \bibinfo{pages}{007}\bibinfo{year}{ (\bibinfo{year}{2006})}}.

\bibitem[{\citenamefont{Amaudruz et~al.}(2018)}]{Amaudruz:2017ekt}
\bibinfo{author}{\bibfnamefont{P.~A.} \bibnamefont{Amaudruz}} \bibnamefont{et~al.},  \href{http://dx.doi.org/10.1103/PhysRevLett.121.071801}{\bibinfo{journal}{Phys. Rev. Lett.} \textbf{\bibinfo{volume}{121}}, \bibinfo{pages}{071801}\bibinfo{year}{ (\bibinfo{year}{2018})}}.

\bibitem[{\citenamefont{Adhikari et~al.}(2021)}]{DEAP:2021axq}
\bibinfo{author}{\bibfnamefont{P.}~\bibnamefont{Adhikari}} \bibnamefont{et~al.},  \href{http://dx.doi.org/10.1140/epjc/s10052-021-09514-w}{\bibinfo{journal}{Eur. Phys. J. C} \textbf{\bibinfo{volume}{81}}, \bibinfo{pages}{823}\bibinfo{year}{ (\bibinfo{year}{2021})}}.

\bibitem[{\citenamefont{Bonivento and Terranova}(2024)}]{Bonivento:2024qpn}
\bibinfo{author}{\bibfnamefont{W.~M.} \bibnamefont{Bonivento}} \bibnamefont{and} \bibinfo{author}{\bibfnamefont{F.}~\bibnamefont{Terranova}},  \href{http://dx.doi.org/10.1103/RevModPhys.96.045001}{\bibinfo{journal}{Rev. Mod. Phys.} \textbf{\bibinfo{volume}{96}}, \bibinfo{pages}{045001}\bibinfo{year}{ (\bibinfo{year}{2024})}}.

\bibitem[{\citenamefont{Acciarri et~al.}(2010)}]{WArP:2008rgv}
\bibinfo{author}{\bibfnamefont{R.}~\bibnamefont{Acciarri}} \bibnamefont{et~al.},  \href{http://dx.doi.org/10.1088/1748-0221/5/06/P06003}{\bibinfo{journal}{JINST} \textbf{\bibinfo{volume}{5}}, \bibinfo{pages}{P06003}\bibinfo{year}{ (\bibinfo{year}{2010})}}.

\bibitem[{\citenamefont{Hartwig}(1994)}]{Hartwig1994}
\bibinfo{author}{\bibfnamefont{G.}~\bibnamefont{Hartwig}}, \emph{\bibinfo{title}{Polymer properties at room and cryogenic temperatures}}, \bibinfo{howpublished}{Plenum Press, New York} (\bibinfo{year}{1994}).

\bibitem[{\citenamefont{Tegeler et~al.}(1999)\citenamefont{Tegeler, Span, and Wagner}}]{10.1063/1.556037}
\bibinfo{author}{\bibfnamefont{C.}~\bibnamefont{Tegeler}}, \bibinfo{author}{\bibfnamefont{R.}~\bibnamefont{Span}},  \bibnamefont{and} \bibinfo{author}{\bibfnamefont{W.}~\bibnamefont{Wagner}},  \href{http://dx.doi.org/10.1063/1.556037}{\bibinfo{journal}{J. Phys. Chem. Ref. Data} \textbf{\bibinfo{volume}{28}}, \bibinfo{pages}{779}\bibinfo{year}{ (\bibinfo{year}{1999})}}.

\bibitem[{\citenamefont{Stewart and Jacobsen}(1989)}]{10.1063/1.555829}
\bibinfo{author}{\bibfnamefont{R.~B.} \bibnamefont{Stewart}} \bibnamefont{and} \bibinfo{author}{\bibfnamefont{R.~T.} \bibnamefont{Jacobsen}},  \href{http://dx.doi.org/10.1063/1.555829}{\bibinfo{journal}{J. Phys. Chem. Ref. Data} \textbf{\bibinfo{volume}{18}}, \bibinfo{pages}{639}\bibinfo{year}{ (\bibinfo{year}{1989})}}.

\bibitem[{\citenamefont{Agnes et~al.}(2017)}]{DarkSide:2016ddo}
\bibinfo{author}{\bibfnamefont{P.}~\bibnamefont{Agnes}} \bibnamefont{et~al.},  \href{http://dx.doi.org/10.1088/1748-0221/12/01/P01021}{\bibinfo{journal}{JINST} \textbf{\bibinfo{volume}{12}}, \bibinfo{pages}{P01021}\bibinfo{year}{ (\bibinfo{year}{2017})}}.

\bibitem[{\citenamefont{Adhikari et~al.}(2025{\natexlab{b}})}]{DEAP:2024mov}
\bibinfo{author}{\bibfnamefont{P.}~\bibnamefont{Adhikari}} \bibnamefont{et~al.},  \href{http://dx.doi.org/10.1140/epjc/s10052-024-13518-7}{\bibinfo{journal}{Eur. Phys. J. C} \textbf{\bibinfo{volume}{85}}, \bibinfo{pages}{87}\bibinfo{year}{ (\bibinfo{year}{2025}{\natexlab{b}})}}.

\bibitem[{\citenamefont{Amaudruz et~al.}(2015)}]{AMAUDRUZ2015178}
\bibinfo{author}{\bibfnamefont{P.-A.} \bibnamefont{Amaudruz}}  \bibnamefont{et~al.},  \href{http://dx.doi.org/10.1016/j.astropartphys.2014.09.006}{\bibinfo{journal}{Astropart. Phys.} \textbf{\bibinfo{volume}{62}}, \bibinfo{pages}{178}\bibinfo{year}{ (\bibinfo{year}{2015})}}.

\bibitem[{\citenamefont{Verkerke and Kirkby}(2003)}]{verkerke2003roofittoolkitdatamodeling}
\bibinfo{author}{\bibfnamefont{W.}~\bibnamefont{Verkerke}} \bibnamefont{and} \bibinfo{author}{\bibfnamefont{D.}~\bibnamefont{Kirkby}}, \emph{\bibinfo{title}{The {RooFit} toolkit for data modeling}} (\bibinfo{year}{2003}), \eprint{physics/0306116}, \urlprefix\url{https://arxiv.org/abs/physics/0306116}.

\end{thebibliography}



\end{document}